\documentclass[prl,superscriptaddress,twocolumn,notitlepage,longbibliography]{revtex4-2}

\usepackage[latin9]{inputenc}
\usepackage{amsfonts}
\usepackage{subfigure}
\usepackage{amsmath}
\usepackage{amssymb}
\usepackage{amsbsy} 
\usepackage{epsfig}
\usepackage{graphicx}
\usepackage{epstopdf}
\usepackage{bbm}
\usepackage[unicode=true,
 bookmarks=true,bookmarksnumbered=false,bookmarksopen=false,
 breaklinks=false,pdfborder={0 0 1},backref=false,colorlinks=true]
 {hyperref}
\hypersetup{
 linkcolor=magenta,urlcolor=blue,citecolor=blue,pdfstartview={FitH},urlcolor=blue}

\def\be{\begin{equation}} \def\ee{\end{equation}}
\def\bea{\begin{eqnarray}} \def\eea{\end{eqnarray}}

\newcommand{\bra}[1]{\langle#1|}
\newcommand{\ket}[1]{|#1\rangle}

\begin{document}
	\title{Non-Bloch dynamics and topology in a classical non-equilibrium process}

\author{Bo Li}
\thanks{These authors contributed equally to this work.}
\affiliation{
Institute for Advanced Study, Tsinghua University, Beijing, 100084, China}

\author{He-Ran Wang}
\thanks{These authors contributed equally to this work.}
\affiliation{
Institute for Advanced Study, Tsinghua University, Beijing, 100084, China}

\author{Fei Song}
\affiliation{
Institute for Advanced Study, Tsinghua University, Beijing, 100084, China}

\author{Zhong Wang}
\email{wangzhongemail@tsinghua.edu.cn}
\affiliation{
Institute for Advanced Study, Tsinghua University, Beijing, 100084, China}
	
\begin{abstract}
The non-Hermitian skin effect refers to the accumulation of eigenstates near the boundary in open boundary lattice models, which can be systematically characterized using the non-Bloch band theory. Here, we apply the non-Bloch band theory to investigate the stochastic reaction-diffusion process by mapping it to a non-Hermitian Kitaev chain. We exactly obtain the open boundary spectrum and the generalized Brillouin zone, and identify a robust zero mode arising from the non-Bloch topology. Notably, distinct from its Hermitian counterpart in the quantum 
context, the zero mode supports anomalous dynamical crossover in the Markov process. We quantitatively demonstrate the intriguing dynamical effects through the spectral decomposition of the Hamiltonian on the non-Bloch eigenstates, and confirm our findings by conducting stochastic simulations with high accuracy. Our study highlights the significant and general role of non-Bloch topology in  non-equilibrium dynamics.  
\end{abstract}

\maketitle
	
The research on stochastic reaction-diffusion processes in one dimension (1D) has been drawing considerable attention in the field of non-equilibrium statistical physics for decades \cite{Evans1993RMPadsoption, Grynberg1994exact,Grynberg1995correlation,Grynberg1995dynamics,Schutz1995Diffusion,Robin2001stochastic,Santos1996,Robertson_2021,Essler_1996}. Despite its simplicity, this model exhibits a wide range of complex behaviors, and has found applications in various disciplines, including physics, chemistry, biology, and even the traffic problems~\cite{Evans1993RMPadsoption,Privman1993dynamics,schadschneider2010stochastic,Volpert2009RD_biology}. The lattice realization of such processes with the exclusion constraint is typically mapped to a fictitious spin system, of which the dynamics is governed by a non-Hermitian many-body Hamiltonian \cite{ALCARAZ1994250,henkel2019reaction}. This intriguing feature makes the model an ideal platform for the rapidly growing study of non-Hermitian physics \cite{Ashida2020,Bergholtz2021RMP}. On the other hand, the recent developments in non-Hermitian physics, especially the remarkable non-Hermitian skin effect (NHSE) \cite{yao2018edge,yao2018chern,kunst2018biorthogonal,
lee2018anatomy,Helbig2019NHSE,Song2019,Longhi2019Probing,alvarez2017,Okuma2020,Zhang2020correspondence,Xiao2020Non,Ghatak2019NHSE,Wang2022defectivetopomode,XiujuanZhang2022reviewNHSE,Yi2020,Xiao2021nonBloch,Xue2022burst} and the interplay between non-Hermiticity and topology \cite{shen2018topological,gong2018nonhermitian, Alvarez2018,leykam2017,kawabata2019symmetry,Borgnia2019,kawabata2019exceptional,liu2019second,Song2019real,Weidemann2020topological,Longhi2020chiral}, shed light on the research of the reaction-diffusion process. 

The fermionic Hamiltonian mapped from the stochastic process generally involves many-body interactions, thus posing challenges for both analytical understanding and effective numerical simulations. However, imposing a certain constraint on the parameters enables a quadratic fermionic description of the model through the Jordan-Wigner transformation \cite{Grynberg1994exact,Grynberg1995correlation,Grynberg1995dynamics}. In our work, we focus on this exactly solvable scenario, where the non-Hermitian free-fermion model features in non-reciprocal hopping and imbalanced pair creation and annihilation. We calculate the complete spectrum under the open boundary condition and the associative generalized Brillouin zone (GBZ) utilizing the non-Bloch band theory \cite{yao2018edge, Yokomizo2019, yokomizo2020non, wang2021nonBloch, Wang2024amoeba}. Furthermore, we identify a robust boundary zero eigenstate in this classical stochastic process, originating from the topology of the non-Bloch band and resembling Majorana zero modes (MZMs) in quantum physics. The existence of the zero mode, along with the extensive many skin modes, have significant implications on the asymptotic dynamics towards the steady state. Our findings are two-fold: first, we show the dramatic change of the damping behaviour between different boundary conditions caused by the NHSE; second, we demonstrate an anomalous crossover of the damping rate between transient and asymptotic dynamics, attributed to the zero mode. Our work provides an example on how non-Hermitian quantum physics manifests itself on the classical level, and in turn unveil novel observable phenomena in the stochastic processes.

\textit{The model.}---We begin by describing the Markov process of exclusive particles in 1D. The particles hop and pair randomly on a chain with $L$ sites, and on each site at most one-particle occupation is permitted. We denote a classical configuration on the chain by $\eta$, $|\eta\rangle=|n_1,n_2,\cdots,n_L\rangle$, and $n_j=0,1$ is the occupation number on the site $j$. We map the $n_j=0~(1)$ state to the spin down (up) to realize a classical-quantum correspondence. For stochastic processes, our focus lies on the probability of observing the configuration $\eta$ at time $t$, i.e. $P_\eta(t)$. We represent the probability distribution by a vector $\ket{P(t)}=\sum_\eta P_\eta(t)\ket{\eta}$. We define the homogeneous distribution by $\ket{I}=\sum_\eta \ket{\eta}$, then the expectation value of observables, e.g. the total particle number $N$, is evaluated by $\sum_\eta N(\eta)P_\eta(t)=\bra{I}N\ket{P(t)}$. The Markovian evolution of the probability distribution is governed by the master equation as $\frac{d}{dt}\ket{P(t)}=-\mathcal{H}\ket{P(t)}$, and the generator $\mathcal{H}$ admits a form of a spin-1/2 Hamiltonian under the classical-to-quantum mapping~\cite{Sandow1994PASEP,Gwa1992sixVertex,Schutz1994NonAbelian,Dhar1987exact}. Real parts of eigenvalues of $\mathcal{H}$ are guaranteed to be non-negative due to the contractive nature of the Markov process, and the zero eigenvalue corresponds to the steady state. Matrix elements of the ``many-body Hamiltonian" $\mathcal{H}$ on the computational bases (eigenstates of the spin-$z$ operators) must be the real number since they account for the transition rates of the Markov process. However, the condition of detailed balance can be broken if we allow for the non-reciprocity transition rates between two configurations: $\mathcal{H}\neq \mathcal{H}^T$, thus leading to the non-Hermiticty.

\begin{figure}
      \includegraphics[width=1\linewidth]{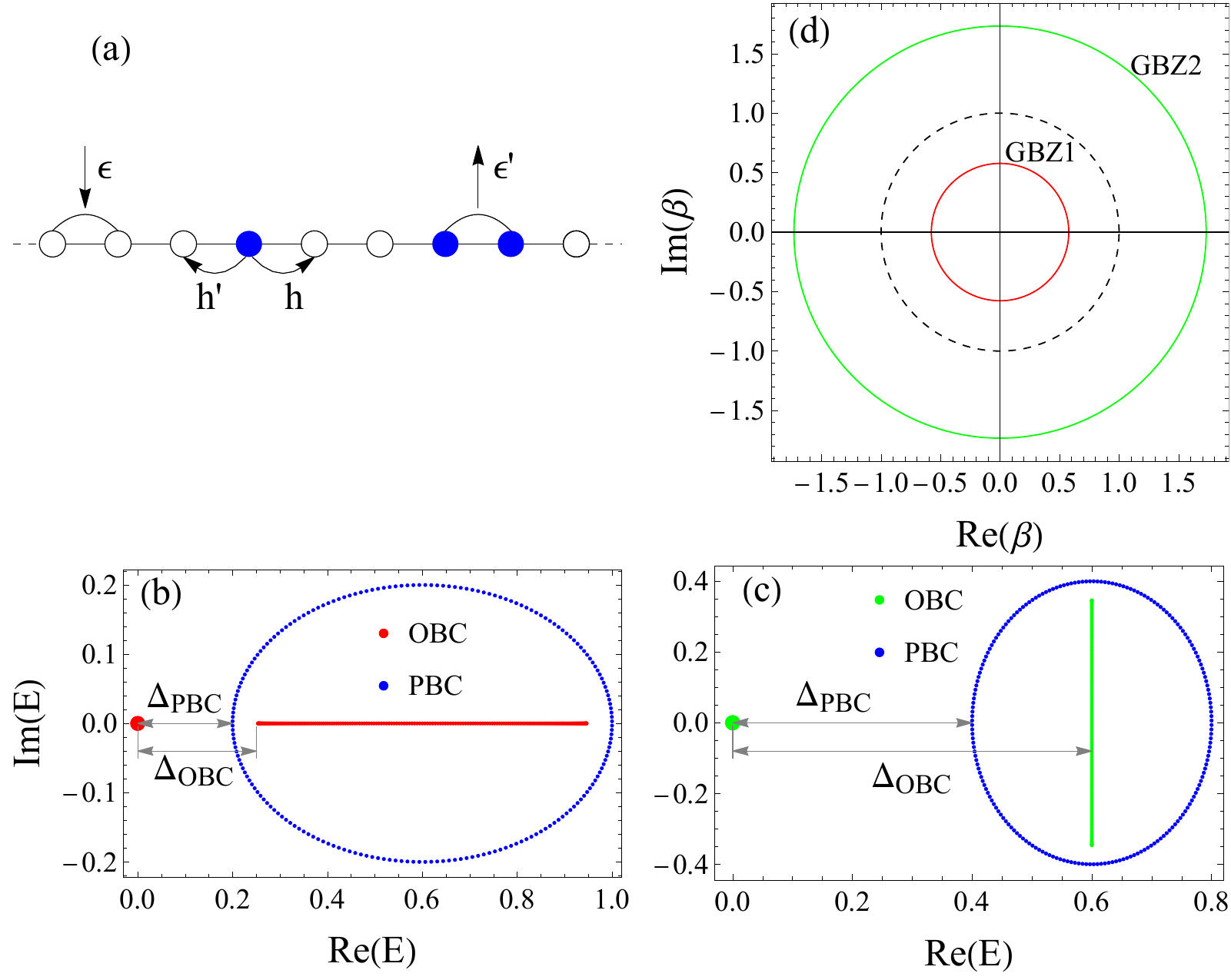}
 \caption{ (a) A schematic plot for the stochastic process. Energy spectrum under different boundary conditions are shown in (b) ($s=0.3, \delta_1=0.2, \delta_2=0.1$) and (c) ($s=0.3, \delta_1=0.1, \delta_2=-0.2$), and energy gaps are marked. (d) The generalized Brillouin zone (GBZ) for the parameters in (b) (red) and (c) (green).}\label{fig:band_GBZ}
\end{figure}

For the specific model, we consider four types of microscopic processes involving adjacent two sites: pair deposition (evaporation) with rate $\epsilon$ ($\epsilon'$), and particle hopping to the right or left with rate $h$ and $h'$ respectively. We represent $\mathcal H$ by a non-Hermitian spin-1/2 Hamiltonian in terms of Pauli matrices $\sigma_j^\alpha$ ($\alpha=x,y,z$) \cite{Schutz1995Diffusion,henkel2019reaction,ALCARAZ1994250,Klobas2023Stochastic}:
\begin{eqnarray*}\label{eqn:model}
\mathcal H=&&\sum_{j}\epsilon[-\hat\sigma_j^+\hat\sigma_{j+1}^++(1-\hat n_j)(1-\hat n_{j+1})]+\epsilon^\prime[-\hat\sigma_j^-\hat\sigma_{j+1}^-\nonumber\\
&&+\hat n_j\hat n_{j+1}]+h[-\hat\sigma_{j+1}^+\hat\sigma_j^-+\hat n_j(1-\hat n_{j+1})]\nonumber\\
&&+h^\prime[-\hat\sigma_j^+\hat\sigma_{j+1}^-+(1-\hat n_{j})\hat n_{j+1}],
\end{eqnarray*}
where the spin flipping operators are $\hat\sigma_j^{\pm}=(\hat\sigma_j^x\pm i\hat\sigma_j^y)/2$, and $\hat n_j=(1+\hat\sigma_j^z)/2=\hat{\sigma}_j^+\hat{\sigma_j}^-$. The index $j$ ranges from $1$ to $L-1$ under the open boundary condition (OBC), while taking $L$ additionally for the periodic boundary condition (PBC). Notice that in the Hamiltonian, each non-diagonal ``$\sigma\sigma$" term is accompanied by a diagonal interacting term to ensure the total probability conservation of the Markov process. We impose the constraint $\epsilon+\epsilon'=h+h'$ for the parameters to make interacting terms proportional to $n_j n_{j+1}$ vanish. Subsequently, we apply the Jordan-Wigner transformation \cite{Jordan1928} on the spin Hamiltonian, leading to the following free-fermion superconducting model:
\begin{eqnarray}\label{eq:NHKitaev}
\mathcal H=&&\sum_{j}  -(\epsilon c_j^\dagger c^\dagger_{j+1}+\epsilon^\prime c_{j+1}c_{j}+h c^\dagger_{j+1}c_j+h^\prime c^\dagger_j c_{j+1})\nonumber\\
&&+(h-\epsilon)c_j^\dagger c_j+(h^\prime-\epsilon)c^\dagger_{j+1}c_{j+1}+\text{const.},
\end{eqnarray}
where $c_j$ ($c_j^\dagger$) is the fermionic annihilation (creation) operator on the site $j$. The effective superconducting pairing originates from the deposition and evaporation processes, as illustrated in Fig. \ref{fig:band_GBZ} (a). 
We stress that the Hamiltonian conserves fermion parity and is always in the ``symmetry breaking" regime: suppose that we start the evolution from an initial state with even/odd parity, it will terminate at a steady state with the same parity after long time, implying the existence of at least two steady states with the opposite parity. The manifestation of the degeneracy differs between PBC and OBC: for the former case, the Jordan-Wigner transformation attaches a parity operator to the hopping and pairing terms involving fermion operators on the site $1$ and $L$,
leading to (anti-) periodic boundary conditions for the fermionic model; on the other hand, under the OBC, the free-fermion Hamiltonian necessarily have an \textit{exact} zero-energy single-particle eigenstate to hold the degeneracy. In that sense, the zero mode is protected by the probability conservation. In the following, we will show that it is also protected by the unique topological properties of the non-Bloch band \footnote{Recently, we became aware of Ref. \cite{Klobas2023Stochastic} which also studied zero modes in stochastic processes, though not in the framework of the non-Bloch band theory, and the physical focus is different.}.
We also note that there exists a broad spectrum of stochastic processes admitting free-fermion representations \cite{Schutz2001Exactly,Supp}, which can all be analyzed harnessing the non-Bloch band theory~\cite{yao2018edge, Yokomizo2019}. Here, we specifically focus on the above model for concreteness.

\textit{The spectrum and the zero mode.--} We adopt the Nambu basis $\Psi=(c_1,\cdots,c_L, c_1^\dagger,\cdots, c_L^\dagger)^T$ to represent the Hamiltonian \eqref{eq:NHKitaev}: $\mathcal H=\frac{1}{2}\Psi^\dagger H\Psi$, where $H$ is a non-Hermitian $2L\times 2L$ matrix satisfying the particle-hole symmetry $\Sigma_x H^T\Sigma_x=-H$ with $\Sigma_x=\sigma^x\bigotimes\mathbbm{1}_L$. The Hamiltonian can be decomposed on its bi-orthogonal eigenstates as $H=\sum_n\varepsilon_n\ket{\psi_n}\bra{\bar{\psi}_n}$, where $\bra{\bar{\psi}_m}\psi_n\rangle=\delta_{mn}$. The particle-hole symmetry guarantees that the eigenvalues appears in pair as $\pm\varepsilon_n$, but in the second-quantized Hamiltonian $\mathcal{H}$ half of them are redundant, such that~\cite{Supp}
\begin{eqnarray}\label{eq:excitation}
\mathcal H=\sum_{\text{Re}(\varepsilon_n)\geq 0,n=1}^L\varepsilon_n\bar{\alpha}^\dagger_n\alpha_n.
\end{eqnarray}
Quasi-particle operators are $\alpha_n=\sum_{l=1}^{2L}\bra{\bar{\psi}_{n}}l\rangle\Psi_l$, $\bar{\alpha}_n=\sum_{l=1}^{2L}\bra{\psi_{n}}l\rangle\Psi_l$, and we assign $n\le L$ to the energies $\varepsilon_n$ with non-negative real part. The quasi-particle operators fulfill the bi-anitcommutation relations $\{\bar{\alpha}_m^\dagger,\alpha_n\}=\delta_{mn}$, inheriting from the bi-orthogonality between eigenstates~\cite{Supp}. The steady state of the Markov process corresponds to the quasi-particle vacuum, and hence the degeneracy can be attributed to the zero eigenvalues of $H$.

We investigate the spectrum of the Hamiltonian $H$ under different boundary conditions to demonstrate the extreme boundary sensitivity of the model. The spectrum under PBC is shown in blue in Fig. \ref{fig:band_GBZ}(b) and (c), forming a closed curve in the complex plane. On the other hand, the OBC spectrum (in red or green) includes a line and an isolated zero mode. For analytical solutions we adopt the parameterization $\epsilon(\epsilon^\prime)=s\pm\delta_1$ 
, $h (h^\prime)=s\pm\delta_2$, where $s>|\delta_{1,2}|$ in the physically relevant regime. We calculate the dispersion relations of the continuum spectrum for both boundary conditions using the non-Bloch band theory \cite{yao2018edge,yokomizo2020non,Supp}:
\begin{eqnarray}\label{eq:spectrum_OBC}
E_{\text{OBC}}(\phi)=&&2s+\sqrt{|\delta_1^2-\delta_2^2|}[\text{sgn}(\delta_1-\delta_2)e^{-i\phi}\nonumber\\
&&+\text{sgn}(\delta_1+\delta_2)e^{i\phi}],
\end{eqnarray}
\begin{eqnarray}
E_{\text{PBC}}(k)=2s+2\delta_1\cos (k)+ 2i\delta_2\sin (k),
\end{eqnarray}
where $\phi$ represents the angle of the complex momentum on the GBZ with the radius $|\beta|=\sqrt{|\delta_1-\delta_2|/|\delta_1+\delta_2|}$, and $k$ is the momentum of eigenstates for PBC [depicted in Fig. \ref{fig:band_GBZ}(d)]. For the isolated zero mode, we can also solve the complex momentum via the non-Bloch band theory: $\beta_{\pm}=(-s\pm\sqrt{s^2-\delta_1^2+\delta_2^2})/(\delta_1+\delta_2)$. From the dispersion relations, we find that the excitation gap between the lowest (real) energy and the zero point is given by $\Delta_{\text{OBC}}=2(s-\text{Re}[\sqrt{\delta_1^2-\delta_2^2}])$ and $\Delta_{\text{PBC}}=2(s-|\delta_1|)$. 

We shall point out that the emergence of the zero mode is intimately related to the non-Bloch topology. Consider a continuous deformation of the Hamiltonian where the parameters $\delta_{1,2}$ are sent to zero while remaining in the physically relevant regime, and the gap $\Delta_{\text{OBC}}$ remains open. In the Hermitian limit with $\delta_{1,2}=0$, the model exactly recovers the 1D Kitaev chain \cite{Kitaev2001Unpaired} which is a well-known example of a topological superconductor hosting unpaired Majorana zero modes localized at the ends of the chain. These Majorana zero modes are described by the operators $\gamma_1=c_1+c_1^\dagger$ and $\gamma_{2L}=i(c_L-c_L^\dagger)$, both of which commute with the Hamiltonian. Therefore, the presence of the zero mode in our non-Hermitian model can be seen as a resemblance to the Majorana zero modes. Furthermore, the zero modes can be diagnosed by the topological invariant on the GBZ~\cite{Kitaev2001Unpaired,Li2023Universal,Supp}. From this perspective, the zero mode is \textit{doubly protected} by the band topology and the probability conservation.

\textit{Non-equilibrium dynamics.--} To show the pronounced consequence on the dynamics brought by the skin modes and the zero mode, we compute the expectation value of the local particle density $\hat{n}_j=c_j^\dagger c_j$ as the system evolves: $\rho_j(t)=\langle I|\hat n_j e^{-\mathcal Ht} |P(0)\rangle$, and we choose the initial state $|P(0)\rangle$ as the empty state without any particle occupation. The expectation value can be divided into a static and a dynamical part as $\rho_j(t)=\rho_{j,\text{st}}+\delta \rho_j(t)$, where the former part represents the static value, and $\delta\rho_j(t)$ captures the time-dependent behavior. We obtain the explicit expression for the two parts \cite{Supp}:
\begin{align}\label{eq:density_st}
\rho_{j,\text{st}}&=\sum_{n=L+1}^{2L}\langle\bar\psi_{n}|j\rangle\langle j|\psi_{n}\rangle,\\
\label{eq:density_dyn}
\delta\rho_j(t)&=\sum_{1\leq n_1<n_2\leq L}\Gamma_{n_1,n_2}^{(j)}\exp[{-(\varepsilon_{n_1}+\varepsilon_{n_2})t}].
\end{align}
Here, $\Gamma_{n_1,n_2}^{(j)}$ depends on the overlap of the two associative eigenstates on the site $j$. Specifically, we set the index $n=1$ for the zero-energy eigenstate, $\varepsilon_1=0$. For indices $n_{1,2}$ in the continuum band, $\Gamma_{n_1,n_2}^{(j)}$ scales as $|\beta|^{2j}$ ($|\beta|=1$ under PBC and $|\beta|=\sqrt{|\delta_1-\delta_2|/|\delta_1+\delta_2|}$ under OBC). On the other hand, when the zero mode is involved, we have $\Gamma_{1,n}^{(j)}\sim |\beta|^j(C_+\beta_+^j+C_-\beta_-^j)$, where $C_{\pm}$ are the coefficients in the zero mode wavefunction.

Having established the unified framework above, we can discuss the discrepancies in the dynamics for different boundary conditions. First, we focus on the asymptotic behavior ($t\gg L$), for which the damping rate $\Lambda$ is determined by the real gap of the many-body Hamiltonian $\mathcal{H}$. Constrained by the parity conservation, $\Lambda$ is given by the two quasi-particle excitations on top of the vacuum (steady state). Under PBC, such the lowest excited state is achieved by filling two quasi-particles near the bottom of the band with the (real) energy $2\Delta_{\text{PBC}}$. However, under OBC we can fill one zero mode and one quasi-particle near the band bottom, resulting in the damping rate 
$\Delta_{\text{OBC}}$. These observations highlight a significant distinction in the dynamics between the two boundary conditions:
\begin{eqnarray}\label{eq:lifetime}
\Lambda_{\text{OBC}}=\Delta_{\text{OBC}},\qquad\Lambda_{\text{PBC}}=2\Delta_{\text{PBC}}.
\end{eqnarray}

\begin{figure}
\begin{tabular}{cc}
      \includegraphics[width=1\linewidth]{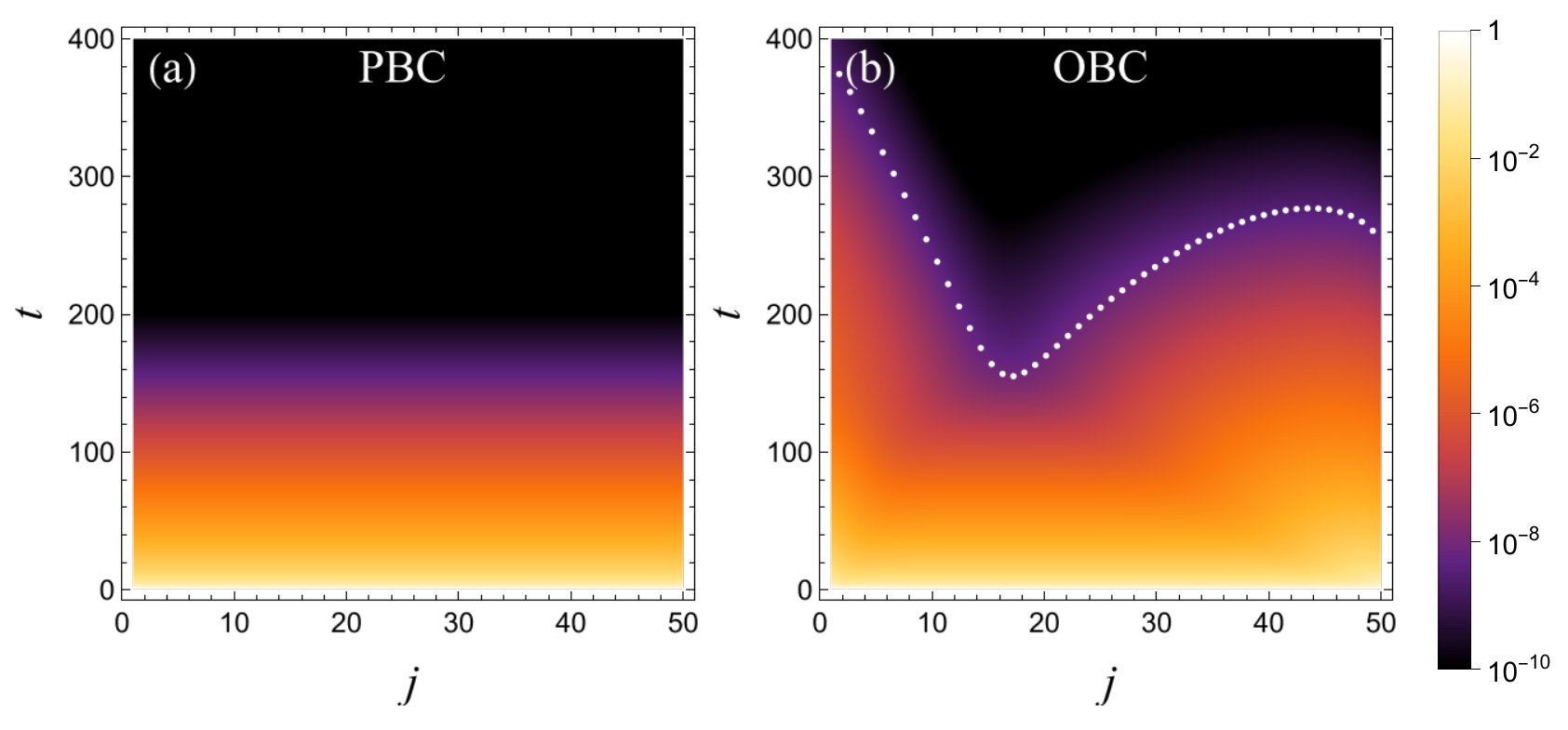}\\
      \includegraphics[width=1\linewidth]{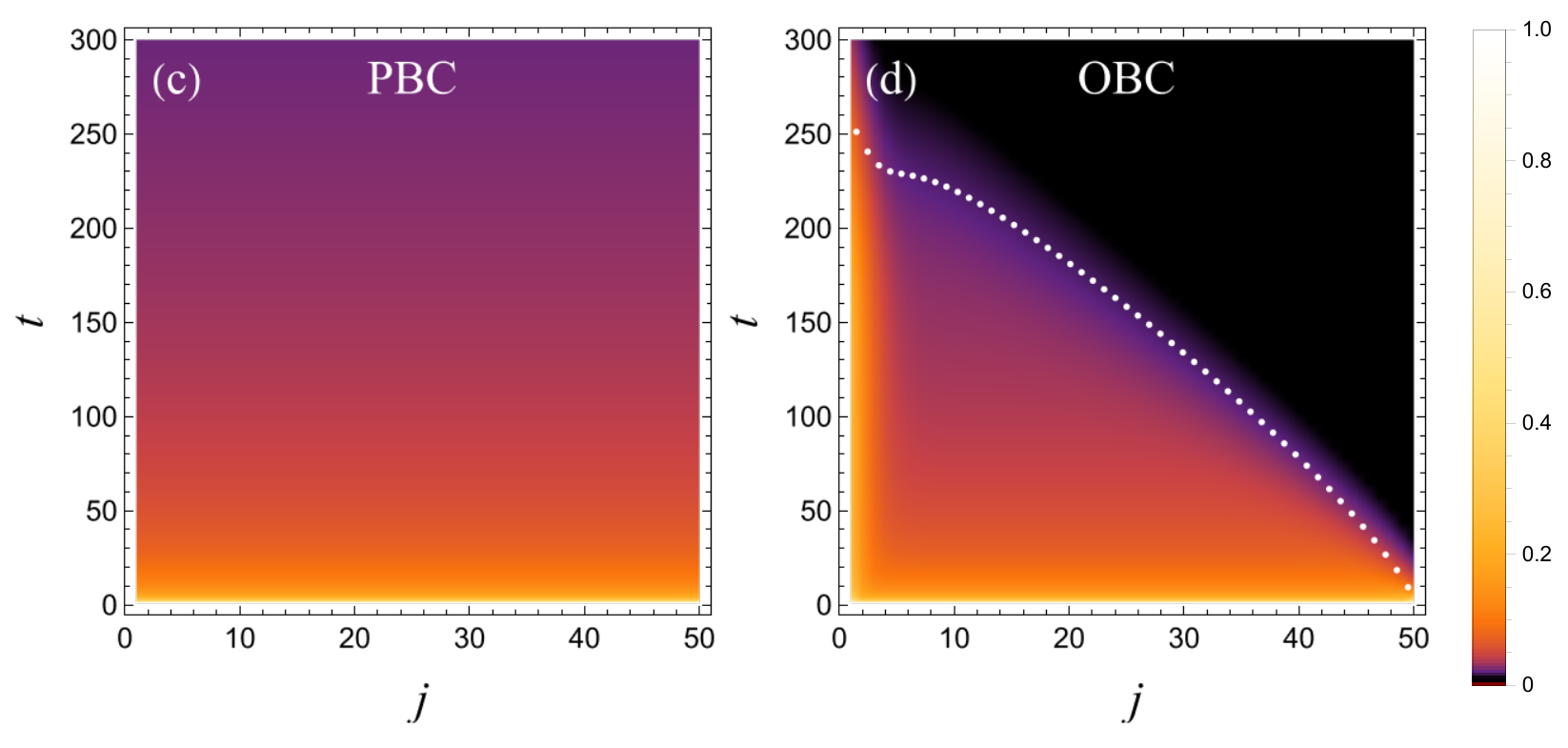}
\end{tabular}
 \caption{The evolution of $|\delta\rho_j(t)|$. In each row, the left and right panel are respectively the evolution under PBC and OBC. The dotted lines in (b) and (d) show $\ln\lambda^{(j)}$ with $t^\ast=400$ and $t^\ast=300$ respectively. Parameters are chosen as: (a)(b) $s=0.3$, $\delta_1=0.28$, $\delta_2=0.05$; (c)(d)  $s=0.3$, $\delta_1=0.3$, $\delta_2=0.1$.}\label{fig:position_evolution}
\end{figure}

Second, we find that the transient dynamics ($t\sim L$) is enriched by the competence between wavefunction overlaps $\Gamma_{n_1,n_2}^{(j)}$, contributed from different eigenstates. To gain insight we again consider the Hermitian limit as an illustrative example, where the zero mode strictly localizes on the two boundary sites $j=1,L$, and the single-particle gap shares the same value between different boundary conditions: $\Delta_{\text{OBC}}=\Delta_{\text{PBC}}=\Delta$. In this case, only on the boundary sites the damping rate equals the one quasi-particle excitation gap. Since the wavefunction of the zero mode has no tail in the bulk, the damping rate should double for generic $j$. Applying Eq. \eqref{eq:density_st} and \eqref{eq:density_dyn} on the Hermitian case, we obtain the damping behavior of particle densities:
\begin{eqnarray}
&&\rho_1(t)=\rho_L(t)=\frac{1}{2}-\frac{1}{2}e^{-\Delta t},\nonumber\\
&&\rho_j(t)=\frac{1}{2}-\frac{1}{2}e^{-2\Delta t}, \qquad (j=2,3,\cdots, L-1).
\end{eqnarray}
This prominent feature of different damping rates between boundary sites and deep in the bulk still holds in the non-Hermitian models, particularly during the transient time, which is imprinted by the presence of the robust zero mode.

In the non-Hermitian cases with $\delta_1,\delta_2\neq 0$, the non-Hermiticity can lead to the penetration of the zero mode into the bulk exponentially. During the initial stage of the evolution, particle densities near the boundaries are quickly influenced by the zero mode and decay with the rate $\Delta_{\text{OBC}}$. In contrast, the particle densities in the bulk do not feel the presence of the boundaries and exhibit a damping rate of $2\Delta_{\text{PBC}}$, until the influence of the zero mode propagates into the bulk \cite{Song2019,Haga2021Liouvillian,Mao2021Boundary}. As the system evolves, densities on all sites will enter the long-time damping stage in sequence depending on the distance to the boundaries, following the same damping rate $\Delta_{\text{OBC}}$.  The site dependence of the transition time between two damping rates readily reflects the profile of the zero mode wavefunction. 

The evolution of $\delta\rho_j(t)$ is depicted in Fig. \ref{fig:position_evolution}, where the sharp color transition indicates that the dynamical part approaches a small value $\kappa$. In Figs. \ref{fig:position_evolution}(a)(b) we choose parameters yielding a finite gap for both OBC and PBC spectrum, thus resulting in exponential decaying dynamics. For the PBC, we expect uniform damping due to the translational invariance, while under OBC the spatially dependent damping arises from the joint effect of the skin modes and the zero mode. To estimate the color transition time $t_c^{(j)}$, we approximate $\delta\rho_j(t)$ by an exponential decay function $\lambda^{(j)}\exp({-\Delta_\text{OBC}t})$, where $\lambda^{(j)}=\sum_{n>1}^L \Gamma^{(j)}_{1,n}\exp[(-\varepsilon_n-\Delta_{\text{OBC}}) t^\ast]$ is evaluated at a characteristic time $t^\ast$. Consequently, the color transition time is well approximated by $\delta\rho_j(t_c)\simeq \kappa$ so that $t_c^{(j)}\simeq \ln{\lambda^{(j)}}/\Delta_{\text{OBC}}+\text{const}.$, implying a coincidence between the spatial dependence of the color transition line and $\ln\lambda^{(j)}$, as proved by the white dotted line in Fig. \ref{fig:position_evolution} (b). In (c)(d), we tune the parameters $s=\delta_1$ to close the gap under PBC, and the results are reminiscent of the chiral damping in open quantum systems \cite{Song2019} (algebraic decaying for PBC while the sharp wavefront for OBC). Remarkably, the special parameters give rise to the delocalization of the zero mode ($|\beta_-|=1$). That explains why $\lambda^{(j)}$ primarily depends on the wavefunction of the NHSE eigenstates, as shown by the dotted line in Figs. \ref{fig:position_evolution}(d).
\begin{figure}
\begin{tabular}{cc}
      \includegraphics[width=1\linewidth]{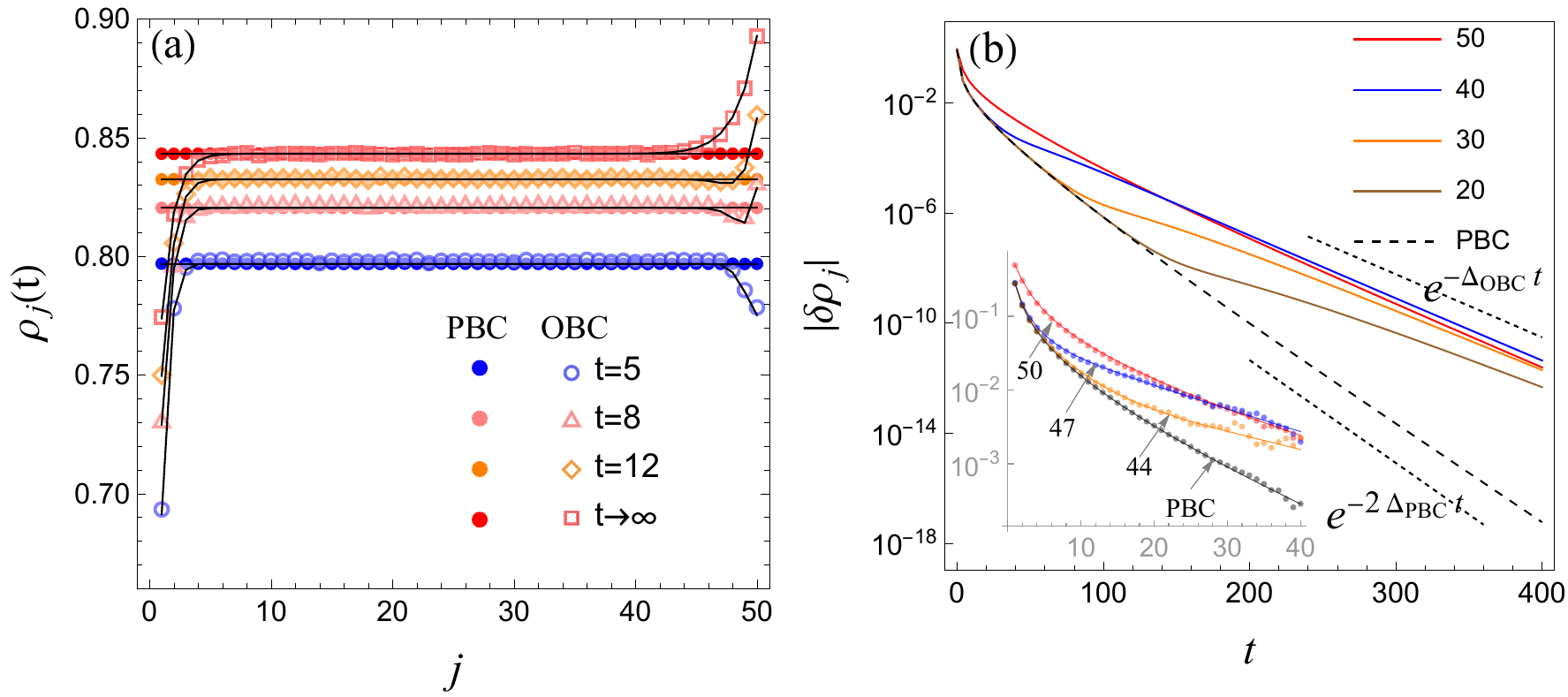}\\
      \includegraphics[width=1\linewidth]{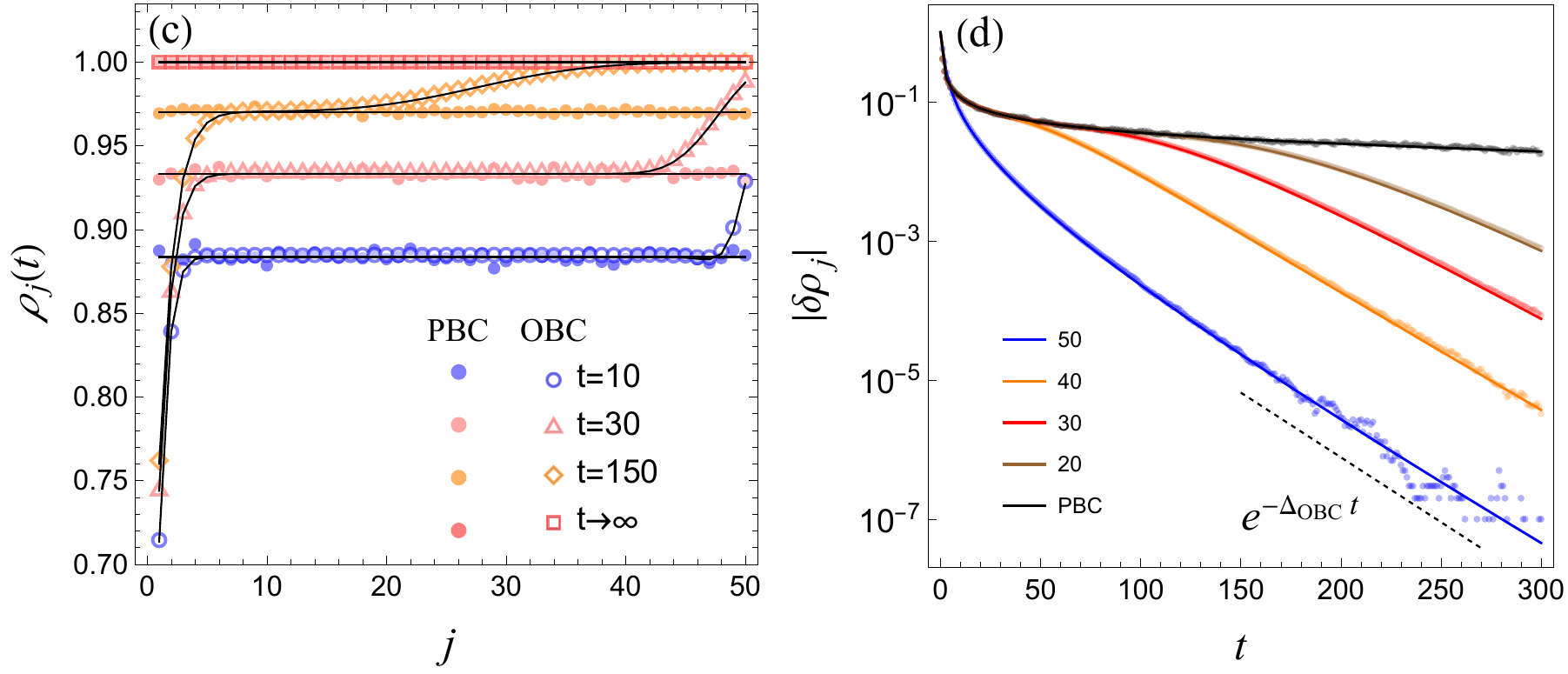}
\end{tabular}
 \caption{(a)(c) Expectation values of particle densities at selected moments, where dots and open markers come from numerical simulations, and black lines are theoretical results. (b)(d) The dynamical part of expectation values at given positions, where dots and lines respectively give numerical and theoretical results.  Parameters are taken as the following: (a)(b) $s=0.3$, $\delta_1=0.28$, $\delta_2=0.05$; (c)(d) $s=0.3$, $\delta_1=0.3$, $\delta_2=0.1$.}\label{fig:density_and_decay}
\end{figure}

In Fig. \ref{fig:density_and_decay}, we zoom in on specific time slices or sites to gain more knowledge about the dynamics. We also incorporate stochastic simulations to validate our theoretical results on the classical-to-quantum mapping \cite{Comment2}. We confirm that under OBC, the evolution on a site in the bulk generally follows the PBC trajectory until it feels the presence of the boundaries. As shown in (a) and (c), the deviation of expectation values between OBC and PBC gradually transmits from the boundaries into the bulk. Additionally, in (b) and (d), we observe sequential departures of the OBC evolution from the PBC trajectory, accompanied by a clear crossover of the damping rates from $2\Delta_{\text{PBC}}$ to $\Delta_{\text{OBC}}$. Notably, in Fig. \ref{fig:density_and_decay}(b), the damping slows down after the crossover due to $2\Delta_{\text{PBC}}>\Delta_{\text{OBC}}$, while in Fig. \ref{fig:density_and_decay}(d) the gapless PBC spectrum leads to an acceleration of the dynamics from algebraic to exponential decaying. This motives us to bring forward the condition for the balanced damping rates in transient and asymptotic dynamics:
\begin{equation}
    2\Delta_{\text{PBC}}=\Delta_{\text{OBC}}\Rightarrow s=2|\delta_1|-\text{Re}[\sqrt{\delta_1^2-\delta_2^2}].
\end{equation}
We emphasize that such a dynamical competition would be impossible in the absence of the zero mode, since then the PBC damping rate cannot be greater than the OBC counterpart because $2\Delta_{\text{OBC}}\ge 2\Delta_{\text{PBC}}$, which is ensured by the non-Bloch band theory \cite{Wang2024amoeba}. Therefore, the phenomenon of dynamical crossover can be regarded as a hallmark of the classical Majorana zero mode.


\textit{Conclusions.--} In summary, we explore the applications of the non-Bloch band theory in a classical stochastic process and uncover the existence of the boundary-localized zero mode, by mapping the Markov generator to a non-Hermitian Kitaev chain. Our work demonstrates the distinct impact of the interplay between the zero mode and skin modes on both transient and asymptotic dynamics, influencing the relaxation time and spatial dependence of the damping behavior. Our discoveries contribute to the understanding of the recent remarkable progress in non-Hermitian physics. Future research directions may include generalizing the classical zero mode to the quantum regime through open quantum dynamics \cite{Bernard2019Open,Essler2020Integrability,Robertson_2021,Bernard2022Dynamics}, and going beyond the current scope with exactly solvable parameters to explore the connections between the zero modes and non-Hermitian interacting topology. 


\textit{Acknowledgements.}--- 
We thank J. Wang, C. Chen, and G.Sun for discussions. This work is supported by the National Natural Science Foundation of China under Grant No.12125405, National Key R\&D Program of China (No. 2023YFA1406702), and the Innovation Program for Quantum Science and Technology (No.2021ZD0302502).
	
\bibliographystyle{apsrev4-1-title}
\bibliography{NHMajorana,dirac}

\begin{thebibliography}{68}%
\makeatletter
\providecommand \@ifxundefined [1]{%
 \@ifx{#1\undefined}
}%
\providecommand \@ifnum [1]{%
 \ifnum #1\expandafter \@firstoftwo
 \else \expandafter \@secondoftwo
 \fi
}%
\providecommand \@ifx [1]{%
 \ifx #1\expandafter \@firstoftwo
 \else \expandafter \@secondoftwo
 \fi
}%
\providecommand \natexlab [1]{#1}%
\providecommand \enquote  [1]{``#1''}%
\providecommand \bibnamefont  [1]{#1}%
\providecommand \bibfnamefont [1]{#1}%
\providecommand \citenamefont [1]{#1}%
\providecommand \href@noop [0]{\@secondoftwo}%
\providecommand \href [0]{\begingroup \@sanitize@url \@href}%
\providecommand \@href[1]{\@@startlink{#1}\@@href}%
\providecommand \@@href[1]{\endgroup#1\@@endlink}%
\providecommand \@sanitize@url [0]{\catcode `\\12\catcode `\$12\catcode
  `\&12\catcode `\#12\catcode `\^12\catcode `\_12\catcode `\%12\relax}%
\providecommand \@@startlink[1]{}%
\providecommand \@@endlink[0]{}%
\providecommand \url  [0]{\begingroup\@sanitize@url \@url }%
\providecommand \@url [1]{\endgroup\@href {#1}{\urlprefix }}%
\providecommand \urlprefix  [0]{URL }%
\providecommand \Eprint [0]{\href }%
\providecommand \doibase [0]{http://dx.doi.org/}%
\providecommand \selectlanguage [0]{\@gobble}%
\providecommand \bibinfo  [0]{\@secondoftwo}%
\providecommand \bibfield  [0]{\@secondoftwo}%
\providecommand \translation [1]{[#1]}%
\providecommand \BibitemOpen [0]{}%
\providecommand \bibitemStop [0]{}%
\providecommand \bibitemNoStop [0]{.\EOS\space}%
\providecommand \EOS [0]{\spacefactor3000\relax}%
\providecommand \BibitemShut  [1]{\csname bibitem#1\endcsname}%
\let\auto@bib@innerbib\@empty
\bibitem [{\citenamefont {Evans}(1993)}]{Evans1993RMPadsoption}%
  \BibitemOpen
  \bibfield  {author} {\bibinfo {author} {\bibfnamefont {J.~W.}\ \bibnamefont
  {Evans}},\ }\bibfield  {title} {\enquote {\bibinfo {title} {Random and
  cooperative sequential adsorption},}\ }\href {\doibase
  10.1103/RevModPhys.65.1281} {\bibfield  {journal} {\bibinfo  {journal} {Rev.
  Mod. Phys.}\ }\textbf {\bibinfo {volume} {65}},\ \bibinfo {pages} {1281}
  (\bibinfo {year} {1993})}\BibitemShut {NoStop}%
\bibitem [{\citenamefont {Grynberg}\ \emph {et~al.}(1994)\citenamefont
  {Grynberg}, \citenamefont {Newman},\ and\ \citenamefont
  {Stinchcombe}}]{Grynberg1994exact}%
  \BibitemOpen
  \bibfield  {author} {\bibinfo {author} {\bibfnamefont {M.~D.}\ \bibnamefont
  {Grynberg}}, \bibinfo {author} {\bibfnamefont {T.~J.}\ \bibnamefont
  {Newman}}, \ and\ \bibinfo {author} {\bibfnamefont {R.~B.}\ \bibnamefont
  {Stinchcombe}},\ }\bibfield  {title} {\enquote {\bibinfo {title} {Exact
  solutions for stochastic adsorption-desorption models and catalytic surface
  processes},}\ }\href {\doibase 10.1103/PhysRevE.50.957} {\bibfield  {journal}
  {\bibinfo  {journal} {Phys. Rev. E}\ }\textbf {\bibinfo {volume} {50}},\
  \bibinfo {pages} {957} (\bibinfo {year} {1994})}\BibitemShut {NoStop}%
\bibitem [{\citenamefont {Grynberg}\ and\ \citenamefont
  {Stinchcombe}(1995{\natexlab{a}})}]{Grynberg1995correlation}%
  \BibitemOpen
  \bibfield  {author} {\bibinfo {author} {\bibfnamefont {M.~D.}\ \bibnamefont
  {Grynberg}}\ and\ \bibinfo {author} {\bibfnamefont {R.~B.}\ \bibnamefont
  {Stinchcombe}},\ }\bibfield  {title} {\enquote {\bibinfo {title} {Dynamic
  correlation functions of adsorption stochastic systems with diffusional
  relaxation},}\ }\href {\doibase 10.1103/PhysRevLett.74.1242} {\bibfield
  {journal} {\bibinfo  {journal} {Phys. Rev. Lett.}\ }\textbf {\bibinfo
  {volume} {74}},\ \bibinfo {pages} {1242} (\bibinfo {year}
  {1995}{\natexlab{a}})}\BibitemShut {NoStop}%
\bibitem [{\citenamefont {Grynberg}\ and\ \citenamefont
  {Stinchcombe}(1995{\natexlab{b}})}]{Grynberg1995dynamics}%
  \BibitemOpen
  \bibfield  {author} {\bibinfo {author} {\bibfnamefont {M.~D.}\ \bibnamefont
  {Grynberg}}\ and\ \bibinfo {author} {\bibfnamefont {R.~B.}\ \bibnamefont
  {Stinchcombe}},\ }\bibfield  {title} {\enquote {\bibinfo {title} {Dynamics of
  adsorption-desorption processes as a soluble problem of many fermions},}\
  }\href {\doibase 10.1103/PhysRevE.52.6013} {\bibfield  {journal} {\bibinfo
  {journal} {Phys. Rev. E}\ }\textbf {\bibinfo {volume} {52}},\ \bibinfo
  {pages} {6013} (\bibinfo {year} {1995}{\natexlab{b}})}\BibitemShut {NoStop}%
\bibitem [{\citenamefont {Schutz}(1995)}]{Schutz1995Diffusion}%
  \BibitemOpen
  \bibfield  {author} {\bibinfo {author} {\bibfnamefont {G.~M.}\ \bibnamefont
  {Schutz}},\ }\bibfield  {title} {\enquote {\bibinfo {title}
  {Diffusion-annihilation in the presence of a driving field},}\ }\href
  {\doibase 10.1088/0305-4470/28/12/014} {\bibfield  {journal} {\bibinfo
  {journal} {, J. Phys. A: Math. Gen.}\ }\textbf {\bibinfo {volume} {28}},\
  \bibinfo {pages} {3405} (\bibinfo {year} {1995})}\BibitemShut {NoStop}%
\bibitem [{\citenamefont {Stinchcombe}(2001)}]{Robin2001stochastic}%
  \BibitemOpen
  \bibfield  {author} {\bibinfo {author} {\bibfnamefont {R.}~\bibnamefont
  {Stinchcombe}},\ }\bibfield  {title} {\enquote {\bibinfo {title} {Stochastic
  non-equilibrium systems},}\ }\href {\doibase 10.1080/00018730110099650}
  {\bibfield  {journal} {\bibinfo  {journal} {Adv. Phys.}\ }\textbf {\bibinfo
  {volume} {50}},\ \bibinfo {pages} {431} (\bibinfo {year} {2001})}\BibitemShut
  {NoStop}%
\bibitem [{\citenamefont {Santos}\ \emph {et~al.}(1996)\citenamefont {Santos},
  \citenamefont {Sch{\"u}tz},\ and\ \citenamefont {Stinchcombe}}]{Santos1996}%
  \BibitemOpen
  \bibfield  {author} {\bibinfo {author} {\bibfnamefont {J.~E.}\ \bibnamefont
  {Santos}}, \bibinfo {author} {\bibfnamefont {G.~M.}\ \bibnamefont
  {Sch{\"u}tz}}, \ and\ \bibinfo {author} {\bibfnamefont {R.~B.}\ \bibnamefont
  {Stinchcombe}},\ }\bibfield  {title} {\enquote {\bibinfo {title}
  {Diffusion--annihilation dynamics in one spatial dimension},}\ }\href
  {\doibase 10.1063/1.472107} {\bibfield  {journal} {\bibinfo  {journal} {J.
  Chem.Phys.}\ }\textbf {\bibinfo {volume} {105}},\ \bibinfo {pages} {2399}
  (\bibinfo {year} {1996})}\BibitemShut {NoStop}%
\bibitem [{\citenamefont {Robertson}\ and\ \citenamefont
  {Essler}(2021)}]{Robertson_2021}%
  \BibitemOpen
  \bibfield  {author} {\bibinfo {author} {\bibfnamefont {J.}~\bibnamefont
  {Robertson}}\ and\ \bibinfo {author} {\bibfnamefont {F.~H.~L.}\ \bibnamefont
  {Essler}},\ }\bibfield  {title} {\enquote {\bibinfo {title} {Exact solution
  of a quantum asymmetric exclusion process with particle creation and
  annihilation},}\ }\href {\doibase 10.1088/1742-5468/ac22f8} {\bibfield
  {journal} {\bibinfo  {journal} {J. Stat. Mech.}\ }\textbf {\bibinfo {volume}
  {2021}},\ \bibinfo {pages} {103102} (\bibinfo {year} {2021})}\BibitemShut
  {NoStop}%
\bibitem [{\citenamefont {Essler}\ and\ \citenamefont
  {Rittenberg}(1996)}]{Essler_1996}%
  \BibitemOpen
  \bibfield  {author} {\bibinfo {author} {\bibfnamefont {F.~H.~L.}\
  \bibnamefont {Essler}}\ and\ \bibinfo {author} {\bibfnamefont
  {V.}~\bibnamefont {Rittenberg}},\ }\bibfield  {title} {\enquote {\bibinfo
  {title} {Representations of the quadratic algebra and partially asymmetric
  diffusion with open boundaries},}\ }\href {\doibase
  10.1088/0305-4470/29/13/013} {\bibfield  {journal} {\bibinfo  {journal} {J.
  Phys. A}\ }\textbf {\bibinfo {volume} {29}},\ \bibinfo {pages} {3375}
  (\bibinfo {year} {1996})}\BibitemShut {NoStop}%
\bibitem [{\citenamefont {Privman}(1994)}]{Privman1993dynamics}%
  \BibitemOpen
  \bibfield  {author} {\bibinfo {author} {\bibfnamefont {V.}~\bibnamefont
  {Privman}},\ }\bibfield  {title} {\enquote {\bibinfo {title} {Dynamics of
  nonequilibrium processes: Surface adsorption, reaction-diffusion kinetics,
  ordering and phase separation},}\ }\href@noop {} {\bibfield  {journal}
  {\bibinfo  {journal} {Trends in Stat. Phys.}\ }\textbf {\bibinfo {volume}
  {1}},\ \bibinfo {pages} {89} (\bibinfo {year} {1994})}\BibitemShut {NoStop}%
\bibitem [{\citenamefont {Schadschneider}\ \emph {et~al.}(2010)\citenamefont
  {Schadschneider}, \citenamefont {Chowdhury},\ and\ \citenamefont
  {Nishinari}}]{schadschneider2010stochastic}%
  \BibitemOpen
  \bibfield  {author} {\bibinfo {author} {\bibfnamefont {A.}~\bibnamefont
  {Schadschneider}}, \bibinfo {author} {\bibfnamefont {D.}~\bibnamefont
  {Chowdhury}}, \ and\ \bibinfo {author} {\bibfnamefont {K.}~\bibnamefont
  {Nishinari}},\ }\href@noop {} {\emph {\bibinfo {title} {Stochastic transport
  in complex systems: from molecules to vehicles}}}\ (\bibinfo  {publisher}
  {Elsevier},\ \bibinfo {year} {2010})\BibitemShut {NoStop}%
\bibitem [{\citenamefont {Volpert}\ and\ \citenamefont
  {Petrovskii}(2009)}]{Volpert2009RD_biology}%
  \BibitemOpen
  \bibfield  {author} {\bibinfo {author} {\bibfnamefont {V.}~\bibnamefont
  {Volpert}}\ and\ \bibinfo {author} {\bibfnamefont {S.}~\bibnamefont
  {Petrovskii}},\ }\bibfield  {title} {\enquote {\bibinfo {title}
  {Reaction-diffusion waves in biology},}\ }\href {\doibase
  https://doi.org/10.1016/j.plrev.2009.10.002} {\bibfield  {journal} {\bibinfo
  {journal} {Phys. Life Rev.}\ }\textbf {\bibinfo {volume} {6}},\ \bibinfo
  {pages} {267} (\bibinfo {year} {2009})}\BibitemShut {NoStop}%
\bibitem [{\citenamefont {Alcaraz}\ \emph {et~al.}(1994)\citenamefont
  {Alcaraz}, \citenamefont {Droz}, \citenamefont {Henkel},\ and\ \citenamefont
  {Rittenberg}}]{ALCARAZ1994250}%
  \BibitemOpen
  \bibfield  {author} {\bibinfo {author} {\bibfnamefont {F.}~\bibnamefont
  {Alcaraz}}, \bibinfo {author} {\bibfnamefont {M.}~\bibnamefont {Droz}},
  \bibinfo {author} {\bibfnamefont {M.}~\bibnamefont {Henkel}}, \ and\ \bibinfo
  {author} {\bibfnamefont {V.}~\bibnamefont {Rittenberg}},\ }\bibfield  {title}
  {\enquote {\bibinfo {title} {Reaction-diffusion processes, critical dynamics,
  and quantum chains},}\ }\href {\doibase
  https://doi.org/10.1006/aphy.1994.1026} {\bibfield  {journal} {\bibinfo
  {journal} {Ann. Phys.}\ }\textbf {\bibinfo {volume} {230}},\ \bibinfo {pages}
  {250} (\bibinfo {year} {1994})}\BibitemShut {NoStop}%
\bibitem [{\citenamefont {Henkel}(2019)}]{henkel2019reaction}%
  \BibitemOpen
  \bibfield  {author} {\bibinfo {author} {\bibfnamefont {M.}~\bibnamefont
  {Henkel}},\ }\href@noop {} {\emph {\bibinfo {title} {Classical and quantum
  nonlinear integrable systems}}}\ (\bibinfo  {publisher} {CRC Press},\
  \bibinfo {year} {2019})\ pp.\ \bibinfo {pages} {256--287}\BibitemShut
  {NoStop}%
\bibitem [{\citenamefont {Ashida}\ \emph {et~al.}(2020)\citenamefont {Ashida},
  \citenamefont {Gong},\ and\ \citenamefont {Ueda}}]{Ashida2020}%
  \BibitemOpen
  \bibfield  {author} {\bibinfo {author} {\bibfnamefont {Y.}~\bibnamefont
  {Ashida}}, \bibinfo {author} {\bibfnamefont {Z.}~\bibnamefont {Gong}}, \ and\
  \bibinfo {author} {\bibfnamefont {M.}~\bibnamefont {Ueda}},\ }\bibfield
  {title} {\enquote {\bibinfo {title} {Non-hermitian physics},}\ }\href
  {\doibase 10.1080/00018732.2021.1876991} {\bibfield  {journal} {\bibinfo
  {journal} {Adv.Phys.}\ }\textbf {\bibinfo {volume} {69}},\ \bibinfo {pages}
  {249} (\bibinfo {year} {2020})}\BibitemShut {NoStop}%
\bibitem [{\citenamefont {Bergholtz}\ \emph {et~al.}(2021)\citenamefont
  {Bergholtz}, \citenamefont {Budich},\ and\ \citenamefont
  {Kunst}}]{Bergholtz2021RMP}%
  \BibitemOpen
  \bibfield  {author} {\bibinfo {author} {\bibfnamefont {E.~J.}\ \bibnamefont
  {Bergholtz}}, \bibinfo {author} {\bibfnamefont {J.~C.}\ \bibnamefont
  {Budich}}, \ and\ \bibinfo {author} {\bibfnamefont {F.~K.}\ \bibnamefont
  {Kunst}},\ }\bibfield  {title} {\enquote {\bibinfo {title} {Exceptional
  topology of non-hermitian systems},}\ }\href {\doibase
  10.1103/RevModPhys.93.015005} {\bibfield  {journal} {\bibinfo  {journal}
  {Rev. Mod. Phys.}\ }\textbf {\bibinfo {volume} {93}},\ \bibinfo {pages}
  {015005} (\bibinfo {year} {2021})}\BibitemShut {NoStop}%
\bibitem [{\citenamefont {Yao}\ and\ \citenamefont {Wang}(2018)}]{yao2018edge}%
  \BibitemOpen
  \bibfield  {author} {\bibinfo {author} {\bibfnamefont {S.}~\bibnamefont
  {Yao}}\ and\ \bibinfo {author} {\bibfnamefont {Z.}~\bibnamefont {Wang}},\
  }\bibfield  {title} {\enquote {\bibinfo {title} {Edge states and topological
  invariants of non-hermitian systems},}\ }\href {\doibase
  10.1103/PhysRevLett.121.086803} {\bibfield  {journal} {\bibinfo  {journal}
  {Phys. Rev. Lett.}\ }\textbf {\bibinfo {volume} {121}},\ \bibinfo {pages}
  {086803} (\bibinfo {year} {2018})}\BibitemShut {NoStop}%
\bibitem [{\citenamefont {Yao}\ \emph {et~al.}(2018)\citenamefont {Yao},
  \citenamefont {Song},\ and\ \citenamefont {Wang}}]{yao2018chern}%
  \BibitemOpen
  \bibfield  {author} {\bibinfo {author} {\bibfnamefont {S.}~\bibnamefont
  {Yao}}, \bibinfo {author} {\bibfnamefont {F.}~\bibnamefont {Song}}, \ and\
  \bibinfo {author} {\bibfnamefont {Z.}~\bibnamefont {Wang}},\ }\bibfield
  {title} {\enquote {\bibinfo {title} {Non-hermitian chern bands},}\ }\href
  {\doibase 10.1103/PhysRevLett.121.136802} {\bibfield  {journal} {\bibinfo
  {journal} {Phys. Rev. Lett.}\ }\textbf {\bibinfo {volume} {121}},\ \bibinfo
  {pages} {136802} (\bibinfo {year} {2018})}\BibitemShut {NoStop}%
\bibitem [{\citenamefont {Kunst}\ \emph {et~al.}(2018)\citenamefont {Kunst},
  \citenamefont {Edvardsson}, \citenamefont {Budich},\ and\ \citenamefont
  {Bergholtz}}]{kunst2018biorthogonal}%
  \BibitemOpen
  \bibfield  {author} {\bibinfo {author} {\bibfnamefont {F.~K.}\ \bibnamefont
  {Kunst}}, \bibinfo {author} {\bibfnamefont {E.}~\bibnamefont {Edvardsson}},
  \bibinfo {author} {\bibfnamefont {J.~C.}\ \bibnamefont {Budich}}, \ and\
  \bibinfo {author} {\bibfnamefont {E.~J.}\ \bibnamefont {Bergholtz}},\
  }\bibfield  {title} {\enquote {\bibinfo {title} {Biorthogonal bulk-boundary
  correspondence in non-hermitian systems},}\ }\href {\doibase
  10.1103/PhysRevLett.121.026808} {\bibfield  {journal} {\bibinfo  {journal}
  {Phys. Rev. Lett.}\ }\textbf {\bibinfo {volume} {121}},\ \bibinfo {pages}
  {026808} (\bibinfo {year} {2018})}\BibitemShut {NoStop}%
\bibitem [{\citenamefont {Lee}\ and\ \citenamefont
  {Thomale}(2019)}]{lee2018anatomy}%
  \BibitemOpen
  \bibfield  {author} {\bibinfo {author} {\bibfnamefont {C.~H.}\ \bibnamefont
  {Lee}}\ and\ \bibinfo {author} {\bibfnamefont {R.}~\bibnamefont {Thomale}},\
  }\bibfield  {title} {\enquote {\bibinfo {title} {Anatomy of skin modes and
  topology in non-hermitian systems},}\ }\href {\doibase
  10.1103/PhysRevB.99.201103} {\bibfield  {journal} {\bibinfo  {journal} {Phys.
  Rev. B}\ }\textbf {\bibinfo {volume} {99}},\ \bibinfo {pages} {201103}
  (\bibinfo {year} {2019})}\BibitemShut {NoStop}%
\bibitem [{\citenamefont {Helbig}\ \emph {et~al.}(2020)\citenamefont {Helbig},
  \citenamefont {Hofmann}, \citenamefont {Imhof}, \citenamefont {Abdelghany},
  \citenamefont {Kiessling}, \citenamefont {Molenkamp}, \citenamefont {Lee},
  \citenamefont {Szameit}, \citenamefont {Greiter},\ and\ \citenamefont
  {Thomale}}]{Helbig2019NHSE}%
  \BibitemOpen
  \bibfield  {author} {\bibinfo {author} {\bibfnamefont {T.}~\bibnamefont
  {Helbig}}, \bibinfo {author} {\bibfnamefont {T.}~\bibnamefont {Hofmann}},
  \bibinfo {author} {\bibfnamefont {S.}~\bibnamefont {Imhof}}, \bibinfo
  {author} {\bibfnamefont {M.}~\bibnamefont {Abdelghany}}, \bibinfo {author}
  {\bibfnamefont {T.}~\bibnamefont {Kiessling}}, \bibinfo {author}
  {\bibfnamefont {L.~W.}\ \bibnamefont {Molenkamp}}, \bibinfo {author}
  {\bibfnamefont {C.~H.}\ \bibnamefont {Lee}}, \bibinfo {author} {\bibfnamefont
  {A.}~\bibnamefont {Szameit}}, \bibinfo {author} {\bibfnamefont
  {M.}~\bibnamefont {Greiter}}, \ and\ \bibinfo {author} {\bibfnamefont
  {R.}~\bibnamefont {Thomale}},\ }\bibfield  {title} {\enquote {\bibinfo
  {title} {Generalized bulk--boundary correspondence in non-hermitian
  topolectrical circuits},}\ }\href {https://doi.org/10.1038/s41567-020-0922-9}
  {\bibfield  {journal} {\bibinfo  {journal} {Nat. Phys.}\ }\textbf {\bibinfo
  {volume} {16}},\ \bibinfo {pages} {747} (\bibinfo {year} {2020})}\BibitemShut
  {NoStop}%
\bibitem [{\citenamefont {Song}\ \emph
  {et~al.}(2019{\natexlab{a}})\citenamefont {Song}, \citenamefont {Yao},\ and\
  \citenamefont {Wang}}]{Song2019}%
  \BibitemOpen
  \bibfield  {author} {\bibinfo {author} {\bibfnamefont {F.}~\bibnamefont
  {Song}}, \bibinfo {author} {\bibfnamefont {S.}~\bibnamefont {Yao}}, \ and\
  \bibinfo {author} {\bibfnamefont {Z.}~\bibnamefont {Wang}},\ }\bibfield
  {title} {\enquote {\bibinfo {title} {Non-hermitian skin effect and chiral
  damping in open quantum systems},}\ }\href {\doibase
  10.1103/PhysRevLett.123.170401} {\bibfield  {journal} {\bibinfo  {journal}
  {Phys. Rev. Lett.}\ }\textbf {\bibinfo {volume} {123}},\ \bibinfo {pages}
  {170401} (\bibinfo {year} {2019}{\natexlab{a}})}\BibitemShut {NoStop}%
\bibitem [{\citenamefont {Longhi}(2019)}]{Longhi2019Probing}%
  \BibitemOpen
  \bibfield  {author} {\bibinfo {author} {\bibfnamefont {S.}~\bibnamefont
  {Longhi}},\ }\bibfield  {title} {\enquote {\bibinfo {title} {Probing
  non-hermitian skin effect and non-bloch phase transitions},}\ }\href
  {\doibase 10.1103/PhysRevResearch.1.023013} {\bibfield  {journal} {\bibinfo
  {journal} {Phys. Rev. Research}\ }\textbf {\bibinfo {volume} {1}},\ \bibinfo
  {pages} {023013} (\bibinfo {year} {2019})}\BibitemShut {NoStop}%
\bibitem [{\citenamefont {Martinez~Alvarez}\ \emph
  {et~al.}(2018{\natexlab{a}})\citenamefont {Martinez~Alvarez}, \citenamefont
  {Barrios~Vargas},\ and\ \citenamefont {Foa~Torres}}]{alvarez2017}%
  \BibitemOpen
  \bibfield  {author} {\bibinfo {author} {\bibfnamefont {V.~M.}\ \bibnamefont
  {Martinez~Alvarez}}, \bibinfo {author} {\bibfnamefont {J.~E.}\ \bibnamefont
  {Barrios~Vargas}}, \ and\ \bibinfo {author} {\bibfnamefont {L.~E.~F.}\
  \bibnamefont {Foa~Torres}},\ }\bibfield  {title} {\enquote {\bibinfo {title}
  {Non-hermitian robust edge states in one dimension: Anomalous localization
  and eigenspace condensation at exceptional points},}\ }\href {\doibase
  10.1103/PhysRevB.97.121401} {\bibfield  {journal} {\bibinfo  {journal} {Phys.
  Rev. B}\ }\textbf {\bibinfo {volume} {97}},\ \bibinfo {pages} {121401}
  (\bibinfo {year} {2018}{\natexlab{a}})}\BibitemShut {NoStop}%
\bibitem [{\citenamefont {Okuma}\ \emph {et~al.}(2020)\citenamefont {Okuma},
  \citenamefont {Kawabata}, \citenamefont {Shiozaki},\ and\ \citenamefont
  {Sato}}]{Okuma2020}%
  \BibitemOpen
  \bibfield  {author} {\bibinfo {author} {\bibfnamefont {N.}~\bibnamefont
  {Okuma}}, \bibinfo {author} {\bibfnamefont {K.}~\bibnamefont {Kawabata}},
  \bibinfo {author} {\bibfnamefont {K.}~\bibnamefont {Shiozaki}}, \ and\
  \bibinfo {author} {\bibfnamefont {M.}~\bibnamefont {Sato}},\ }\bibfield
  {title} {\enquote {\bibinfo {title} {Topological origin of non-hermitian skin
  effects},}\ }\href {\doibase 10.1103/PhysRevLett.124.086801} {\bibfield
  {journal} {\bibinfo  {journal} {Phys. Rev. Lett.}\ }\textbf {\bibinfo
  {volume} {124}},\ \bibinfo {pages} {086801} (\bibinfo {year}
  {2020})}\BibitemShut {NoStop}%
\bibitem [{\citenamefont {Zhang}\ \emph {et~al.}(2020)\citenamefont {Zhang},
  \citenamefont {Yang},\ and\ \citenamefont {Fang}}]{Zhang2020correspondence}%
  \BibitemOpen
  \bibfield  {author} {\bibinfo {author} {\bibfnamefont {K.}~\bibnamefont
  {Zhang}}, \bibinfo {author} {\bibfnamefont {Z.}~\bibnamefont {Yang}}, \ and\
  \bibinfo {author} {\bibfnamefont {C.}~\bibnamefont {Fang}},\ }\bibfield
  {title} {\enquote {\bibinfo {title} {Correspondence between winding numbers
  and skin modes in non-hermitian systems},}\ }\href {\doibase
  10.1103/PhysRevLett.125.126402} {\bibfield  {journal} {\bibinfo  {journal}
  {Phys. Rev. Lett.}\ }\textbf {\bibinfo {volume} {125}},\ \bibinfo {pages}
  {126402} (\bibinfo {year} {2020})}\BibitemShut {NoStop}%
\bibitem [{\citenamefont {Xiao}\ \emph {et~al.}(2020)\citenamefont {Xiao},
  \citenamefont {Deng}, \citenamefont {Wang}, \citenamefont {Zhu},
  \citenamefont {Wang}, \citenamefont {Yi},\ and\ \citenamefont
  {Xue}}]{Xiao2020Non}%
  \BibitemOpen
  \bibfield  {author} {\bibinfo {author} {\bibfnamefont {L.}~\bibnamefont
  {Xiao}}, \bibinfo {author} {\bibfnamefont {T.}~\bibnamefont {Deng}}, \bibinfo
  {author} {\bibfnamefont {K.}~\bibnamefont {Wang}}, \bibinfo {author}
  {\bibfnamefont {G.}~\bibnamefont {Zhu}}, \bibinfo {author} {\bibfnamefont
  {Z.}~\bibnamefont {Wang}}, \bibinfo {author} {\bibfnamefont {W.}~\bibnamefont
  {Yi}}, \ and\ \bibinfo {author} {\bibfnamefont {P.}~\bibnamefont {Xue}},\
  }\bibfield  {title} {\enquote {\bibinfo {title} {Non-hermitian bulk--boundary
  correspondence in quantum dynamics},}\ }\href
  {https://doi.org/10.1038/s41567-020-0836-6} {\bibfield  {journal} {\bibinfo
  {journal} {Nat. Phys.}\ }\textbf {\bibinfo {volume} {16}},\ \bibinfo {pages}
  {761} (\bibinfo {year} {2020})}\BibitemShut {NoStop}%
\bibitem [{\citenamefont {Ghatak}\ \emph {et~al.}(2020)\citenamefont {Ghatak},
  \citenamefont {Brandenbourger}, \citenamefont {van Wezel},\ and\
  \citenamefont {Coulais}}]{Ghatak2019NHSE}%
  \BibitemOpen
  \bibfield  {author} {\bibinfo {author} {\bibfnamefont {A.}~\bibnamefont
  {Ghatak}}, \bibinfo {author} {\bibfnamefont {M.}~\bibnamefont
  {Brandenbourger}}, \bibinfo {author} {\bibfnamefont {J.}~\bibnamefont {van
  Wezel}}, \ and\ \bibinfo {author} {\bibfnamefont {C.}~\bibnamefont
  {Coulais}},\ }\bibfield  {title} {\enquote {\bibinfo {title} {Observation of
  non-hermitian topology and its bulk--edge correspondence in an active
  mechanical metamaterial},}\ }\href {https://doi.org/10.1073/pnas.2010580117}
  {\bibfield  {journal} {\bibinfo  {journal} {Proc. Natl. Acad. Sci. U.S.A.}\
  }\textbf {\bibinfo {volume} {117}},\ \bibinfo {pages} {29561} (\bibinfo
  {year} {2020})}\BibitemShut {NoStop}%
\bibitem [{\citenamefont {Wang}\ \emph {et~al.}(2022)\citenamefont {Wang},
  \citenamefont {Wang},\ and\ \citenamefont {Ma}}]{Wang2022defectivetopomode}%
  \BibitemOpen
  \bibfield  {author} {\bibinfo {author} {\bibfnamefont {W.}~\bibnamefont
  {Wang}}, \bibinfo {author} {\bibfnamefont {X.}~\bibnamefont {Wang}}, \ and\
  \bibinfo {author} {\bibfnamefont {G.}~\bibnamefont {Ma}},\ }\bibfield
  {title} {\enquote {\bibinfo {title} {Non-hermitian morphing of topological
  modes},}\ }\href {\doibase 10.1038/s41586-022-04929-1} {\bibfield  {journal}
  {\bibinfo  {journal} {Nature}\ }\textbf {\bibinfo {volume} {608}},\ \bibinfo
  {pages} {50} (\bibinfo {year} {2022})}\BibitemShut {NoStop}%
\bibitem [{\citenamefont {Zhang}\ \emph {et~al.}(2022)\citenamefont {Zhang},
  \citenamefont {Zhang}, \citenamefont {Lu},\ and\ \citenamefont
  {Chen}}]{XiujuanZhang2022reviewNHSE}%
  \BibitemOpen
  \bibfield  {author} {\bibinfo {author} {\bibfnamefont {X.}~\bibnamefont
  {Zhang}}, \bibinfo {author} {\bibfnamefont {T.}~\bibnamefont {Zhang}},
  \bibinfo {author} {\bibfnamefont {M.-H.}\ \bibnamefont {Lu}}, \ and\ \bibinfo
  {author} {\bibfnamefont {Y.-F.}\ \bibnamefont {Chen}},\ }\bibfield  {title}
  {\enquote {\bibinfo {title} {A review on non-hermitian skin effect},}\ }\href
  {\doibase 10.1080/23746149.2022.2109431} {\bibfield  {journal} {\bibinfo
  {journal} {Adv.Phys.: X}\ }\textbf {\bibinfo {volume} {7}},\ \bibinfo {pages}
  {2109431} (\bibinfo {year} {2022})}\BibitemShut {NoStop}%
\bibitem [{\citenamefont {Yi}\ and\ \citenamefont {Yang}(2020)}]{Yi2020}%
  \BibitemOpen
  \bibfield  {author} {\bibinfo {author} {\bibfnamefont {Y.}~\bibnamefont
  {Yi}}\ and\ \bibinfo {author} {\bibfnamefont {Z.}~\bibnamefont {Yang}},\
  }\bibfield  {title} {\enquote {\bibinfo {title} {Non-hermitian skin modes
  induced by on-site dissipations and chiral tunneling effect},}\ }\href
  {\doibase 10.1103/PhysRevLett.125.186802} {\bibfield  {journal} {\bibinfo
  {journal} {Phys. Rev. Lett.}\ }\textbf {\bibinfo {volume} {125}},\ \bibinfo
  {pages} {186802} (\bibinfo {year} {2020})}\BibitemShut {NoStop}%
\bibitem [{\citenamefont {Xiao}\ \emph {et~al.}(2021)\citenamefont {Xiao},
  \citenamefont {Deng}, \citenamefont {Wang}, \citenamefont {Wang},
  \citenamefont {Yi},\ and\ \citenamefont {Xue}}]{Xiao2021nonBloch}%
  \BibitemOpen
  \bibfield  {author} {\bibinfo {author} {\bibfnamefont {L.}~\bibnamefont
  {Xiao}}, \bibinfo {author} {\bibfnamefont {T.}~\bibnamefont {Deng}}, \bibinfo
  {author} {\bibfnamefont {K.}~\bibnamefont {Wang}}, \bibinfo {author}
  {\bibfnamefont {Z.}~\bibnamefont {Wang}}, \bibinfo {author} {\bibfnamefont
  {W.}~\bibnamefont {Yi}}, \ and\ \bibinfo {author} {\bibfnamefont
  {P.}~\bibnamefont {Xue}},\ }\bibfield  {title} {\enquote {\bibinfo {title}
  {Observation of non-bloch parity-time symmetry and exceptional points},}\
  }\href {\doibase 10.1103/PhysRevLett.126.230402} {\bibfield  {journal}
  {\bibinfo  {journal} {Phys. Rev. Lett.}\ }\textbf {\bibinfo {volume} {126}},\
  \bibinfo {pages} {230402} (\bibinfo {year} {2021})}\BibitemShut {NoStop}%
\bibitem [{\citenamefont {Xue}\ \emph {et~al.}(2022)\citenamefont {Xue},
  \citenamefont {Hu}, \citenamefont {Song},\ and\ \citenamefont
  {Wang}}]{Xue2022burst}%
  \BibitemOpen
  \bibfield  {author} {\bibinfo {author} {\bibfnamefont {W.-T.}\ \bibnamefont
  {Xue}}, \bibinfo {author} {\bibfnamefont {Y.-M.}\ \bibnamefont {Hu}},
  \bibinfo {author} {\bibfnamefont {F.}~\bibnamefont {Song}}, \ and\ \bibinfo
  {author} {\bibfnamefont {Z.}~\bibnamefont {Wang}},\ }\bibfield  {title}
  {\enquote {\bibinfo {title} {Non-hermitian edge burst},}\ }\href {\doibase
  10.1103/PhysRevLett.128.120401} {\bibfield  {journal} {\bibinfo  {journal}
  {Phys. Rev. Lett.}\ }\textbf {\bibinfo {volume} {128}},\ \bibinfo {pages}
  {120401} (\bibinfo {year} {2022})}\BibitemShut {NoStop}%
\bibitem [{\citenamefont {Shen}\ \emph {et~al.}(2018)\citenamefont {Shen},
  \citenamefont {Zhen},\ and\ \citenamefont {Fu}}]{shen2018topological}%
  \BibitemOpen
  \bibfield  {author} {\bibinfo {author} {\bibfnamefont {H.}~\bibnamefont
  {Shen}}, \bibinfo {author} {\bibfnamefont {B.}~\bibnamefont {Zhen}}, \ and\
  \bibinfo {author} {\bibfnamefont {L.}~\bibnamefont {Fu}},\ }\bibfield
  {title} {\enquote {\bibinfo {title} {Topological band theory for
  non-hermitian hamiltonians},}\ }\href {\doibase
  10.1103/PhysRevLett.120.146402} {\bibfield  {journal} {\bibinfo  {journal}
  {Phys. Rev. Lett.}\ }\textbf {\bibinfo {volume} {120}},\ \bibinfo {pages}
  {146402} (\bibinfo {year} {2018})}\BibitemShut {NoStop}%
\bibitem [{\citenamefont {Gong}\ \emph {et~al.}(2018)\citenamefont {Gong},
  \citenamefont {Ashida}, \citenamefont {Kawabata}, \citenamefont {Takasan},
  \citenamefont {Higashikawa},\ and\ \citenamefont
  {Ueda}}]{gong2018nonhermitian}%
  \BibitemOpen
  \bibfield  {author} {\bibinfo {author} {\bibfnamefont {Z.}~\bibnamefont
  {Gong}}, \bibinfo {author} {\bibfnamefont {Y.}~\bibnamefont {Ashida}},
  \bibinfo {author} {\bibfnamefont {K.}~\bibnamefont {Kawabata}}, \bibinfo
  {author} {\bibfnamefont {K.}~\bibnamefont {Takasan}}, \bibinfo {author}
  {\bibfnamefont {S.}~\bibnamefont {Higashikawa}}, \ and\ \bibinfo {author}
  {\bibfnamefont {M.}~\bibnamefont {Ueda}},\ }\bibfield  {title} {\enquote
  {\bibinfo {title} {Topological phases of non-hermitian systems},}\ }\href
  {\doibase 10.1103/PhysRevX.8.031079} {\bibfield  {journal} {\bibinfo
  {journal} {Phys. Rev. X}\ }\textbf {\bibinfo {volume} {8}},\ \bibinfo {pages}
  {031079} (\bibinfo {year} {2018})}\BibitemShut {NoStop}%
\bibitem [{\citenamefont {Martinez~Alvarez}\ \emph
  {et~al.}(2018{\natexlab{b}})\citenamefont {Martinez~Alvarez}, \citenamefont
  {Barrios~Vargas}, \citenamefont {Berdakin},\ and\ \citenamefont
  {Foa~Torres}}]{Alvarez2018}%
  \BibitemOpen
  \bibfield  {author} {\bibinfo {author} {\bibfnamefont {V.~M.}\ \bibnamefont
  {Martinez~Alvarez}}, \bibinfo {author} {\bibfnamefont {J.~E.}\ \bibnamefont
  {Barrios~Vargas}}, \bibinfo {author} {\bibfnamefont {M.}~\bibnamefont
  {Berdakin}}, \ and\ \bibinfo {author} {\bibfnamefont {L.~E.~F.}\ \bibnamefont
  {Foa~Torres}},\ }\bibfield  {title} {\enquote {\bibinfo {title} {Topological
  states of non-hermitian systems},}\ }\href {\doibase
  10.1140/epjst/e2018-800091-5} {\bibfield  {journal} {\bibinfo  {journal}
  {Eur. Phys. J. Spec. Top.}\ }\textbf {\bibinfo {volume} {227}},\ \bibinfo
  {pages} {1295} (\bibinfo {year} {2018}{\natexlab{b}})}\BibitemShut {NoStop}%
\bibitem [{\citenamefont {Leykam}\ \emph {et~al.}(2017)\citenamefont {Leykam},
  \citenamefont {Bliokh}, \citenamefont {Huang}, \citenamefont {Chong},\ and\
  \citenamefont {Nori}}]{leykam2017}%
  \BibitemOpen
  \bibfield  {author} {\bibinfo {author} {\bibfnamefont {D.}~\bibnamefont
  {Leykam}}, \bibinfo {author} {\bibfnamefont {K.~Y.}\ \bibnamefont {Bliokh}},
  \bibinfo {author} {\bibfnamefont {C.}~\bibnamefont {Huang}}, \bibinfo
  {author} {\bibfnamefont {Y.~D.}\ \bibnamefont {Chong}}, \ and\ \bibinfo
  {author} {\bibfnamefont {F.}~\bibnamefont {Nori}},\ }\bibfield  {title}
  {\enquote {\bibinfo {title} {Edge modes, degeneracies, and topological
  numbers in non-hermitian systems},}\ }\href {\doibase
  10.1103/PhysRevLett.118.040401} {\bibfield  {journal} {\bibinfo  {journal}
  {Phys. Rev. Lett.}\ }\textbf {\bibinfo {volume} {118}},\ \bibinfo {pages}
  {040401} (\bibinfo {year} {2017})}\BibitemShut {NoStop}%
\bibitem [{\citenamefont {Kawabata}\ \emph
  {et~al.}(2019{\natexlab{a}})\citenamefont {Kawabata}, \citenamefont
  {Shiozaki}, \citenamefont {Ueda},\ and\ \citenamefont
  {Sato}}]{kawabata2019symmetry}%
  \BibitemOpen
  \bibfield  {author} {\bibinfo {author} {\bibfnamefont {K.}~\bibnamefont
  {Kawabata}}, \bibinfo {author} {\bibfnamefont {K.}~\bibnamefont {Shiozaki}},
  \bibinfo {author} {\bibfnamefont {M.}~\bibnamefont {Ueda}}, \ and\ \bibinfo
  {author} {\bibfnamefont {M.}~\bibnamefont {Sato}},\ }\bibfield  {title}
  {\enquote {\bibinfo {title} {Symmetry and topology in non-hermitian
  physics},}\ }\href {\doibase 10.1103/PhysRevX.9.041015} {\bibfield  {journal}
  {\bibinfo  {journal} {Phys. Rev. X}\ }\textbf {\bibinfo {volume} {9}},\
  \bibinfo {pages} {041015} (\bibinfo {year} {2019}{\natexlab{a}})}\BibitemShut
  {NoStop}%
\bibitem [{\citenamefont {Borgnia}\ \emph {et~al.}(2020)\citenamefont
  {Borgnia}, \citenamefont {Kruchkov},\ and\ \citenamefont
  {Slager}}]{Borgnia2019}%
  \BibitemOpen
  \bibfield  {author} {\bibinfo {author} {\bibfnamefont {D.~S.}\ \bibnamefont
  {Borgnia}}, \bibinfo {author} {\bibfnamefont {A.~J.}\ \bibnamefont
  {Kruchkov}}, \ and\ \bibinfo {author} {\bibfnamefont {R.-J.}\ \bibnamefont
  {Slager}},\ }\bibfield  {title} {\enquote {\bibinfo {title} {Non-hermitian
  boundary modes and topology},}\ }\href {\doibase
  10.1103/PhysRevLett.124.056802} {\bibfield  {journal} {\bibinfo  {journal}
  {Phys. Rev. Lett.}\ }\textbf {\bibinfo {volume} {124}},\ \bibinfo {pages}
  {056802} (\bibinfo {year} {2020})}\BibitemShut {NoStop}%
\bibitem [{\citenamefont {Kawabata}\ \emph
  {et~al.}(2019{\natexlab{b}})\citenamefont {Kawabata}, \citenamefont
  {Bessho},\ and\ \citenamefont {Sato}}]{kawabata2019exceptional}%
  \BibitemOpen
  \bibfield  {author} {\bibinfo {author} {\bibfnamefont {K.}~\bibnamefont
  {Kawabata}}, \bibinfo {author} {\bibfnamefont {T.}~\bibnamefont {Bessho}}, \
  and\ \bibinfo {author} {\bibfnamefont {M.}~\bibnamefont {Sato}},\ }\bibfield
  {title} {\enquote {\bibinfo {title} {Classification of exceptional points and
  non-hermitian topological semimetals},}\ }\href {\doibase
  10.1103/PhysRevLett.123.066405} {\bibfield  {journal} {\bibinfo  {journal}
  {Phys. Rev. Lett.}\ }\textbf {\bibinfo {volume} {123}},\ \bibinfo {pages}
  {066405} (\bibinfo {year} {2019}{\natexlab{b}})}\BibitemShut {NoStop}%
\bibitem [{\citenamefont {Liu}\ \emph {et~al.}(2019)\citenamefont {Liu},
  \citenamefont {Zhang}, \citenamefont {Ai}, \citenamefont {Gong},
  \citenamefont {Kawabata}, \citenamefont {Ueda},\ and\ \citenamefont
  {Nori}}]{liu2019second}%
  \BibitemOpen
  \bibfield  {author} {\bibinfo {author} {\bibfnamefont {T.}~\bibnamefont
  {Liu}}, \bibinfo {author} {\bibfnamefont {Y.-R.}\ \bibnamefont {Zhang}},
  \bibinfo {author} {\bibfnamefont {Q.}~\bibnamefont {Ai}}, \bibinfo {author}
  {\bibfnamefont {Z.}~\bibnamefont {Gong}}, \bibinfo {author} {\bibfnamefont
  {K.}~\bibnamefont {Kawabata}}, \bibinfo {author} {\bibfnamefont
  {M.}~\bibnamefont {Ueda}}, \ and\ \bibinfo {author} {\bibfnamefont
  {F.}~\bibnamefont {Nori}},\ }\bibfield  {title} {\enquote {\bibinfo {title}
  {Second-order topological phases in non-hermitian systems},}\ }\href
  {\doibase 10.1103/PhysRevLett.122.076801} {\bibfield  {journal} {\bibinfo
  {journal} {Phys. Rev. Lett.}\ }\textbf {\bibinfo {volume} {122}},\ \bibinfo
  {pages} {076801} (\bibinfo {year} {2019})}\BibitemShut {NoStop}%
\bibitem [{\citenamefont {Song}\ \emph
  {et~al.}(2019{\natexlab{b}})\citenamefont {Song}, \citenamefont {Yao},\ and\
  \citenamefont {Wang}}]{Song2019real}%
  \BibitemOpen
  \bibfield  {author} {\bibinfo {author} {\bibfnamefont {F.}~\bibnamefont
  {Song}}, \bibinfo {author} {\bibfnamefont {S.}~\bibnamefont {Yao}}, \ and\
  \bibinfo {author} {\bibfnamefont {Z.}~\bibnamefont {Wang}},\ }\bibfield
  {title} {\enquote {\bibinfo {title} {Non-hermitian topological invariants in
  real space},}\ }\href {\doibase 10.1103/PhysRevLett.123.246801} {\bibfield
  {journal} {\bibinfo  {journal} {Phys. Rev. Lett.}\ }\textbf {\bibinfo
  {volume} {123}},\ \bibinfo {pages} {246801} (\bibinfo {year}
  {2019}{\natexlab{b}})}\BibitemShut {NoStop}%
\bibitem [{\citenamefont {Weidemann}\ \emph {et~al.}(2020)\citenamefont
  {Weidemann}, \citenamefont {Kremer}, \citenamefont {Helbig}, \citenamefont
  {Hofmann}, \citenamefont {Stegmaier}, \citenamefont {Greiter}, \citenamefont
  {Thomale},\ and\ \citenamefont {Szameit}}]{Weidemann2020topological}%
  \BibitemOpen
  \bibfield  {author} {\bibinfo {author} {\bibfnamefont {S.}~\bibnamefont
  {Weidemann}}, \bibinfo {author} {\bibfnamefont {M.}~\bibnamefont {Kremer}},
  \bibinfo {author} {\bibfnamefont {T.}~\bibnamefont {Helbig}}, \bibinfo
  {author} {\bibfnamefont {T.}~\bibnamefont {Hofmann}}, \bibinfo {author}
  {\bibfnamefont {A.}~\bibnamefont {Stegmaier}}, \bibinfo {author}
  {\bibfnamefont {M.}~\bibnamefont {Greiter}}, \bibinfo {author} {\bibfnamefont
  {R.}~\bibnamefont {Thomale}}, \ and\ \bibinfo {author} {\bibfnamefont
  {A.}~\bibnamefont {Szameit}},\ }\bibfield  {title} {\enquote {\bibinfo
  {title} {Topological funneling of light},}\ }\href@noop {} {\bibfield
  {journal} {\bibinfo  {journal} {Science}\ }\textbf {\bibinfo {volume}
  {368}},\ \bibinfo {pages} {311} (\bibinfo {year} {2020})}\BibitemShut
  {NoStop}%
\bibitem [{\citenamefont {Longhi}(2020)}]{Longhi2020chiral}%
  \BibitemOpen
  \bibfield  {author} {\bibinfo {author} {\bibfnamefont {S.}~\bibnamefont
  {Longhi}},\ }\bibfield  {title} {\enquote {\bibinfo {title} {Non-bloch-band
  collapse and chiral zener tunneling},}\ }\href {\doibase
  10.1103/PhysRevLett.124.066602} {\bibfield  {journal} {\bibinfo  {journal}
  {Phys. Rev. Lett.}\ }\textbf {\bibinfo {volume} {124}},\ \bibinfo {pages}
  {066602} (\bibinfo {year} {2020})}\BibitemShut {NoStop}%
\bibitem [{\citenamefont {Yokomizo}\ and\ \citenamefont
  {Murakami}(2019)}]{Yokomizo2019}%
  \BibitemOpen
  \bibfield  {author} {\bibinfo {author} {\bibfnamefont {K.}~\bibnamefont
  {Yokomizo}}\ and\ \bibinfo {author} {\bibfnamefont {S.}~\bibnamefont
  {Murakami}},\ }\bibfield  {title} {\enquote {\bibinfo {title} {Non-bloch band
  theory of non-hermitian systems},}\ }\href {\doibase
  10.1103/PhysRevLett.123.066404} {\bibfield  {journal} {\bibinfo  {journal}
  {Phys. Rev. Lett.}\ }\textbf {\bibinfo {volume} {123}},\ \bibinfo {pages}
  {066404} (\bibinfo {year} {2019})}\BibitemShut {NoStop}%
\bibitem [{\citenamefont {Yokomizo}\ and\ \citenamefont
  {Murakami}(2020)}]{yokomizo2020non}%
  \BibitemOpen
  \bibfield  {author} {\bibinfo {author} {\bibfnamefont {K.}~\bibnamefont
  {Yokomizo}}\ and\ \bibinfo {author} {\bibfnamefont {S.}~\bibnamefont
  {Murakami}},\ }\bibfield  {title} {\enquote {\bibinfo {title} {Non-bloch band
  theory and bulk--edge correspondence in non-hermitian systems},}\ }\href
  {https://doi.org/10.1093/ptep/ptaa140} {\bibfield  {journal} {\bibinfo
  {journal} {Prog. Theor. Exp. Phys.}\ }\textbf {\bibinfo {volume} {2020}},\
  \bibinfo {pages} {12A102} (\bibinfo {year} {2020})}\BibitemShut {NoStop}%
\bibitem [{\citenamefont {Wang}(2021)}]{wang2021nonBloch}%
  \BibitemOpen
  \bibfield  {author} {\bibinfo {author} {\bibfnamefont {Z.}~\bibnamefont
  {Wang}},\ }\bibfield  {title} {\enquote {\bibinfo {title} {Non-bloch band
  theory and beyond},}\ }in\ \href {https://doi.org/10.1142/9789811231711_0016}
  {\emph {\bibinfo {booktitle} {Memorial Volume for Shoucheng Zhang}}}\
  (\bibinfo  {publisher} {World Scientific},\ \bibinfo {year} {2021})\ pp.\
  \bibinfo {pages} {365--387}\BibitemShut {NoStop}%
\bibitem [{\citenamefont {Wang}\ \emph {et~al.}(2024)\citenamefont {Wang},
  \citenamefont {Song},\ and\ \citenamefont {Wang}}]{Wang2024amoeba}%
  \BibitemOpen
  \bibfield  {author} {\bibinfo {author} {\bibfnamefont {H.-Y.}\ \bibnamefont
  {Wang}}, \bibinfo {author} {\bibfnamefont {F.}~\bibnamefont {Song}}, \ and\
  \bibinfo {author} {\bibfnamefont {Z.}~\bibnamefont {Wang}},\ }\bibfield
  {title} {\enquote {\bibinfo {title} {Amoeba formulation of non-bloch band
  theory in arbitrary dimensions},}\ }\href {\doibase
  10.1103/PhysRevX.14.021011} {\bibfield  {journal} {\bibinfo  {journal} {Phys.
  Rev. X}\ }\textbf {\bibinfo {volume} {14}},\ \bibinfo {pages} {021011}
  (\bibinfo {year} {2024})}\BibitemShut {NoStop}%
\bibitem [{\citenamefont {Sandow}(1994)}]{Sandow1994PASEP}%
  \BibitemOpen
  \bibfield  {author} {\bibinfo {author} {\bibfnamefont {S.}~\bibnamefont
  {Sandow}},\ }\bibfield  {title} {\enquote {\bibinfo {title} {Partially
  asymmetric exclusion process with open boundaries},}\ }\href {\doibase
  10.1103/PhysRevE.50.2660} {\bibfield  {journal} {\bibinfo  {journal} {Phys.
  Rev. E}\ }\textbf {\bibinfo {volume} {50}},\ \bibinfo {pages} {2660}
  (\bibinfo {year} {1994})}\BibitemShut {NoStop}%
\bibitem [{\citenamefont {Gwa}\ and\ \citenamefont
  {Spohn}(1992)}]{Gwa1992sixVertex}%
  \BibitemOpen
  \bibfield  {author} {\bibinfo {author} {\bibfnamefont {L.-H.}\ \bibnamefont
  {Gwa}}\ and\ \bibinfo {author} {\bibfnamefont {H.}~\bibnamefont {Spohn}},\
  }\bibfield  {title} {\enquote {\bibinfo {title} {Six-vertex model, roughened
  surfaces, and an asymmetric spin hamiltonian},}\ }\href {\doibase
  10.1103/PhysRevLett.68.725} {\bibfield  {journal} {\bibinfo  {journal} {Phys.
  Rev. Lett.}\ }\textbf {\bibinfo {volume} {68}},\ \bibinfo {pages} {725}
  (\bibinfo {year} {1992})}\BibitemShut {NoStop}%
\bibitem [{\citenamefont {Sch\"utz}\ and\ \citenamefont
  {Sandow}(1994)}]{Schutz1994NonAbelian}%
  \BibitemOpen
  \bibfield  {author} {\bibinfo {author} {\bibfnamefont {G.}~\bibnamefont
  {Sch\"utz}}\ and\ \bibinfo {author} {\bibfnamefont {S.}~\bibnamefont
  {Sandow}},\ }\bibfield  {title} {\enquote {\bibinfo {title} {Non-abelian
  symmetries of stochastic processes: Derivation of correlation functions for
  random-vertex models and disordered-interacting-particle systems},}\ }\href
  {\doibase 10.1103/PhysRevE.49.2726} {\bibfield  {journal} {\bibinfo
  {journal} {Phys. Rev. E}\ }\textbf {\bibinfo {volume} {49}},\ \bibinfo
  {pages} {2726} (\bibinfo {year} {1994})}\BibitemShut {NoStop}%
\bibitem [{\citenamefont {Dhar}(1987)}]{Dhar1987exact}%
  \BibitemOpen
  \bibfield  {author} {\bibinfo {author} {\bibfnamefont {D.}~\bibnamefont
  {Dhar}},\ }\bibfield  {title} {\enquote {\bibinfo {title} {An exactly solved
  model for interfacial growth},}\ }\href {\doibase 10.1080/01411598708241334}
  {\bibfield  {journal} {\bibinfo  {journal} {Phase Transitions}\ }\textbf
  {\bibinfo {volume} {9}},\ \bibinfo {pages} {51} (\bibinfo {year}
  {1987})}\BibitemShut {NoStop}%
\bibitem [{\citenamefont {Klobas}\ \emph {et~al.}(2023)\citenamefont {Klobas},
  \citenamefont {Fendley},\ and\ \citenamefont
  {Garrahan}}]{Klobas2023Stochastic}%
  \BibitemOpen
  \bibfield  {author} {\bibinfo {author} {\bibfnamefont {K.}~\bibnamefont
  {Klobas}}, \bibinfo {author} {\bibfnamefont {P.}~\bibnamefont {Fendley}}, \
  and\ \bibinfo {author} {\bibfnamefont {J.~P.}\ \bibnamefont {Garrahan}},\
  }\bibfield  {title} {\enquote {\bibinfo {title} {Stochastic strong zero modes
  and their dynamical manifestations},}\ }\href {\doibase
  10.1103/PhysRevE.107.L042104} {\bibfield  {journal} {\bibinfo  {journal}
  {Phys. Rev. E}\ }\textbf {\bibinfo {volume} {107}},\ \bibinfo {pages}
  {L042104} (\bibinfo {year} {2023})}\BibitemShut {NoStop}%
\bibitem [{\citenamefont {Wigner}\ and\ \citenamefont
  {Jordan}(1928)}]{Jordan1928}%
  \BibitemOpen
  \bibfield  {author} {\bibinfo {author} {\bibfnamefont {E.}~\bibnamefont
  {Wigner}}\ and\ \bibinfo {author} {\bibfnamefont {P.}~\bibnamefont
  {Jordan}},\ }\bibfield  {title} {\enquote {\bibinfo {title} {{\"U}ber das
  paulische {\"a}quivalenzverbot},}\ }\href
  {https://doi.org/10.1007/BF01331938} {\bibfield  {journal} {\bibinfo
  {journal} {Z. Phys}\ }\textbf {\bibinfo {volume} {47}},\ \bibinfo {pages}
  {631} (\bibinfo {year} {1928})}\BibitemShut {NoStop}%
\bibitem [{Note1()}]{Note1}%
  \BibitemOpen
  \bibinfo {note} {Recently, we became aware of Ref. \cite
  {Klobas2023Stochastic} which also studied zero modes in stochastic processes,
  though not in the framework of the non-Bloch band theory, and the physical
  focus is different.}\BibitemShut {Stop}%
\bibitem [{\citenamefont {Sch$\ddot{\text{u}}$tz}(2001)}]{Schutz2001Exactly}%
  \BibitemOpen
  \bibfield  {author} {\bibinfo {author} {\bibfnamefont {G.}~\bibnamefont
  {Sch$\ddot{\text{u}}$tz}},\ }\bibfield  {title} {\enquote {\bibinfo {title}
  {Exactly solvable models for many-body systems far from equilibrium},}\ \
  }(\bibinfo  {publisher} {Elsevier},\ \bibinfo {year} {2001})\ pp.\ \bibinfo
  {pages} {1--251}\BibitemShut {NoStop}%
\bibitem [{Sup()}]{Supp}%
  \BibitemOpen
  \href@noop {} {}\bibinfo {note} {See the Supplementary Materials at [URL will
  be inserted by publisher] for details of calculations (see also references
  \cite{Hatano1996,Kawabata2020Symplectic_nonBloch,Li2020critical}
  therein).}\BibitemShut {Stop}%
\bibitem [{\citenamefont {Kitaev}(2001)}]{Kitaev2001Unpaired}%
  \BibitemOpen
  \bibfield  {author} {\bibinfo {author} {\bibfnamefont {A.~Y.}\ \bibnamefont
  {Kitaev}},\ }\bibfield  {title} {\enquote {\bibinfo {title} {Unpaired
  majorana fermions in quantum wires},}\ }\href {\doibase
  10.1070/1063-7869/44/10S/S29} {\bibfield  {journal} {\bibinfo  {journal}
  {Phys.-Usp.}\ }\textbf {\bibinfo {volume} {44}},\ \bibinfo {pages} {131}
  (\bibinfo {year} {2001})}\BibitemShut {NoStop}%
\bibitem [{\citenamefont {Li}\ \emph {et~al.}(2022)\citenamefont {Li},
  \citenamefont {Cao}, \citenamefont {Chen},\ and\ \citenamefont
  {Yang}}]{Li2023Universal}%
  \BibitemOpen
  \bibfield  {author} {\bibinfo {author} {\bibfnamefont {Y.}~\bibnamefont
  {Li}}, \bibinfo {author} {\bibfnamefont {Y.}~\bibnamefont {Cao}}, \bibinfo
  {author} {\bibfnamefont {Y.}~\bibnamefont {Chen}}, \ and\ \bibinfo {author}
  {\bibfnamefont {X.}~\bibnamefont {Yang}},\ }\bibfield  {title} {\enquote
  {\bibinfo {title} {Universal characteristics of one-dimensional non-hermitian
  superconductors},}\ }\href {\doibase 10.1088/1361-648X/aca4b4} {\bibfield
  {journal} {\bibinfo  {journal} {J. Phys.: Condens. Matter}\ }\textbf
  {\bibinfo {volume} {35}},\ \bibinfo {pages} {055401} (\bibinfo {year}
  {2022})}\BibitemShut {NoStop}%
\bibitem [{\citenamefont {Haga}\ \emph {et~al.}(2021)\citenamefont {Haga},
  \citenamefont {Nakagawa}, \citenamefont {Hamazaki},\ and\ \citenamefont
  {Ueda}}]{Haga2021Liouvillian}%
  \BibitemOpen
  \bibfield  {author} {\bibinfo {author} {\bibfnamefont {T.}~\bibnamefont
  {Haga}}, \bibinfo {author} {\bibfnamefont {M.}~\bibnamefont {Nakagawa}},
  \bibinfo {author} {\bibfnamefont {R.}~\bibnamefont {Hamazaki}}, \ and\
  \bibinfo {author} {\bibfnamefont {M.}~\bibnamefont {Ueda}},\ }\bibfield
  {title} {\enquote {\bibinfo {title} {Liouvillian skin effect: Slowing down of
  relaxation processes without gap closing},}\ }\href {\doibase
  10.1103/PhysRevLett.127.070402} {\bibfield  {journal} {\bibinfo  {journal}
  {Phys. Rev. Lett.}\ }\textbf {\bibinfo {volume} {127}},\ \bibinfo {pages}
  {070402} (\bibinfo {year} {2021})}\BibitemShut {NoStop}%
\bibitem [{\citenamefont {Mao}\ \emph {et~al.}(2021)\citenamefont {Mao},
  \citenamefont {Deng},\ and\ \citenamefont {Zhang}}]{Mao2021Boundary}%
  \BibitemOpen
  \bibfield  {author} {\bibinfo {author} {\bibfnamefont {L.}~\bibnamefont
  {Mao}}, \bibinfo {author} {\bibfnamefont {T.}~\bibnamefont {Deng}}, \ and\
  \bibinfo {author} {\bibfnamefont {P.}~\bibnamefont {Zhang}},\ }\bibfield
  {title} {\enquote {\bibinfo {title} {Boundary condition independence of
  non-hermitian hamiltonian dynamics},}\ }\href {\doibase
  10.1103/PhysRevB.104.125435} {\bibfield  {journal} {\bibinfo  {journal}
  {Phys. Rev. B}\ }\textbf {\bibinfo {volume} {104}},\ \bibinfo {pages}
  {125435} (\bibinfo {year} {2021})}\BibitemShut {NoStop}%
\bibitem [{Com()}]{Comment2}%
  \BibitemOpen
  \href@noop {} {}\bibinfo {note} {The simulation procedure involves
  numerically implementing the stochastic processes described in the main text
  by following Ref.~\cite{Grynberg1995dynamics}. We identify $L$ steps as one
  unit of time. The simulation is stopped at a certain step to complete one
  trajectory, and we repeat this process for $N$ times to average over all
  trajectories. The precision is controlled by $N$, e.g., the simulation data
  presented in Fig. 3 (d) is generated with $N=10^7$, and the simulation
  deviates from the theoretical results as $\sim N^{-1}$. Notice that short
  plateaus can be observed in $\delta\rho_{50}$ around $t=200\sim 300$ since
  the averaged expectation value has already reached the steady state within
  the precision. In Fig. 3 (b), the simulation is only performed up to 40 units
  of time (see the inset) due to limitations in precision}\BibitemShut
  {NoStop}%
\bibitem [{\citenamefont {Bernard}\ and\ \citenamefont
  {Jin}(2019)}]{Bernard2019Open}%
  \BibitemOpen
  \bibfield  {author} {\bibinfo {author} {\bibfnamefont {D.}~\bibnamefont
  {Bernard}}\ and\ \bibinfo {author} {\bibfnamefont {T.}~\bibnamefont {Jin}},\
  }\bibfield  {title} {\enquote {\bibinfo {title} {Open quantum symmetric
  simple exclusion process},}\ }\href {\doibase 10.1103/PhysRevLett.123.080601}
  {\bibfield  {journal} {\bibinfo  {journal} {Phys. Rev. Lett.}\ }\textbf
  {\bibinfo {volume} {123}},\ \bibinfo {pages} {080601} (\bibinfo {year}
  {2019})}\BibitemShut {NoStop}%
\bibitem [{\citenamefont {Essler}\ and\ \citenamefont
  {Piroli}(2020)}]{Essler2020Integrability}%
  \BibitemOpen
  \bibfield  {author} {\bibinfo {author} {\bibfnamefont {F.~H.~L.}\
  \bibnamefont {Essler}}\ and\ \bibinfo {author} {\bibfnamefont
  {L.}~\bibnamefont {Piroli}},\ }\bibfield  {title} {\enquote {\bibinfo {title}
  {Integrability of one-dimensional lindbladians from operator-space
  fragmentation},}\ }\href {\doibase 10.1103/PhysRevE.102.062210} {\bibfield
  {journal} {\bibinfo  {journal} {Phys. Rev. E}\ }\textbf {\bibinfo {volume}
  {102}},\ \bibinfo {pages} {062210} (\bibinfo {year} {2020})}\BibitemShut
  {NoStop}%
\bibitem [{\citenamefont {Bernard}\ \emph {et~al.}(2022)\citenamefont
  {Bernard}, \citenamefont {Essler}, \citenamefont {Hruza},\ and\ \citenamefont
  {Medenjak}}]{Bernard2022Dynamics}%
  \BibitemOpen
  \bibfield  {author} {\bibinfo {author} {\bibfnamefont {D.}~\bibnamefont
  {Bernard}}, \bibinfo {author} {\bibfnamefont {F.}~\bibnamefont {Essler}},
  \bibinfo {author} {\bibfnamefont {L.}~\bibnamefont {Hruza}}, \ and\ \bibinfo
  {author} {\bibfnamefont {M.}~\bibnamefont {Medenjak}},\ }\bibfield  {title}
  {\enquote {\bibinfo {title} {Dynamics of fluctuations in quantum simple
  exclusion processes},}\ }\href {\doibase 10.21468/SciPostPhys.12.1.042}
  {\bibfield  {journal} {\bibinfo  {journal} {SciPost Phys.}\ }\textbf
  {\bibinfo {volume} {12}},\ \bibinfo {pages} {042} (\bibinfo {year}
  {2022})}\BibitemShut {NoStop}%
\bibitem [{\citenamefont {Hatano}\ and\ \citenamefont
  {Nelson}(1996)}]{Hatano1996}%
  \BibitemOpen
  \bibfield  {author} {\bibinfo {author} {\bibfnamefont {N.}~\bibnamefont
  {Hatano}}\ and\ \bibinfo {author} {\bibfnamefont {D.~R.}\ \bibnamefont
  {Nelson}},\ }\bibfield  {title} {\enquote {\bibinfo {title} {Localization
  transitions in non-hermitian quantum mechanics},}\ }\href {\doibase
  10.1103/PhysRevLett.77.570} {\bibfield  {journal} {\bibinfo  {journal} {Phys.
  Rev. Lett.}\ }\textbf {\bibinfo {volume} {77}},\ \bibinfo {pages} {570}
  (\bibinfo {year} {1996})}\BibitemShut {NoStop}%
\bibitem [{\citenamefont {Kawabata}\ \emph {et~al.}(2020)\citenamefont
  {Kawabata}, \citenamefont {Okuma},\ and\ \citenamefont
  {Sato}}]{Kawabata2020Symplectic_nonBloch}%
  \BibitemOpen
  \bibfield  {author} {\bibinfo {author} {\bibfnamefont {K.}~\bibnamefont
  {Kawabata}}, \bibinfo {author} {\bibfnamefont {N.}~\bibnamefont {Okuma}}, \
  and\ \bibinfo {author} {\bibfnamefont {M.}~\bibnamefont {Sato}},\ }\bibfield
  {title} {\enquote {\bibinfo {title} {Non-bloch band theory of non-hermitian
  hamiltonians in the symplectic class},}\ }\href {\doibase
  10.1103/PhysRevB.101.195147} {\bibfield  {journal} {\bibinfo  {journal}
  {Phys. Rev. B}\ }\textbf {\bibinfo {volume} {101}},\ \bibinfo {pages}
  {195147} (\bibinfo {year} {2020})}\BibitemShut {NoStop}%
\bibitem [{\citenamefont {{Li}}\ \emph {et~al.}(2020)\citenamefont {{Li}},
  \citenamefont {{Lee}}, \citenamefont {{Mu}},\ and\ \citenamefont
  {{Gong}}}]{Li2020critical}%
  \BibitemOpen
  \bibfield  {author} {\bibinfo {author} {\bibfnamefont {L.}~\bibnamefont
  {{Li}}}, \bibinfo {author} {\bibfnamefont {C.~H.}\ \bibnamefont {{Lee}}},
  \bibinfo {author} {\bibfnamefont {S.}~\bibnamefont {{Mu}}}, \ and\ \bibinfo
  {author} {\bibfnamefont {J.}~\bibnamefont {{Gong}}},\ }\bibfield  {title}
  {\enquote {\bibinfo {title} {{Critical non-Hermitian skin effect}},}\ }\href
  {\doibase 10.1038/s41467-020-18917-4} {\bibfield  {journal} {\bibinfo
  {journal} {Nat.Commun.}\ }\textbf {\bibinfo {volume} {11}},\ \bibinfo {eid}
  {5491} (\bibinfo {year} {2020})}\BibitemShut {NoStop}%
\end{thebibliography}%


\begin{thebibliography}{9}%
\makeatletter
\providecommand \@ifxundefined [1]{%
 \@ifx{#1\undefined}
}%
\providecommand \@ifnum [1]{%
 \ifnum #1\expandafter \@firstoftwo
 \else \expandafter \@secondoftwo
 \fi
}%
\providecommand \@ifx [1]{%
 \ifx #1\expandafter \@firstoftwo
 \else \expandafter \@secondoftwo
 \fi
}%
\providecommand \natexlab [1]{#1}%
\providecommand \enquote  [1]{``#1''}%
\providecommand \bibnamefont  [1]{#1}%
\providecommand \bibfnamefont [1]{#1}%
\providecommand \citenamefont [1]{#1}%
\providecommand \href@noop [0]{\@secondoftwo}%
\providecommand \href [0]{\begingroup \@sanitize@url \@href}%
\providecommand \@href[1]{\@@startlink{#1}\@@href}%
\providecommand \@@href[1]{\endgroup#1\@@endlink}%
\providecommand \@sanitize@url [0]{\catcode `\\12\catcode `\$12\catcode
  `\&12\catcode `\#12\catcode `\^12\catcode `\_12\catcode `\%12\relax}%
\providecommand \@@startlink[1]{}%
\providecommand \@@endlink[0]{}%
\providecommand \url  [0]{\begingroup\@sanitize@url \@url }%
\providecommand \@url [1]{\endgroup\@href {#1}{\urlprefix }}%
\providecommand \urlprefix  [0]{URL }%
\providecommand \Eprint [0]{\href }%
\providecommand \doibase [0]{https://doi.org/}%
\providecommand \selectlanguage [0]{\@gobble}%
\providecommand \bibinfo  [0]{\@secondoftwo}%
\providecommand \bibfield  [0]{\@secondoftwo}%
\providecommand \translation [1]{[#1]}%
\providecommand \BibitemOpen [0]{}%
\providecommand \bibitemStop [0]{}%
\providecommand \bibitemNoStop [0]{.\EOS\space}%
\providecommand \EOS [0]{\spacefactor3000\relax}%
\providecommand \BibitemShut  [1]{\csname bibitem#1\endcsname}%
\let\auto@bib@innerbib\@empty
\bibitem [{\citenamefont {Sch$\ddot{\text{u}}$tz}(2001)}]{Schutz2001Exactly}%
  \BibitemOpen
  \bibfield  {author} {\bibinfo {author} {\bibfnamefont {G.}~\bibnamefont
  {Sch$\ddot{\text{u}}$tz}},\ }\bibfield  {title} {\bibinfo {title} {Exactly
  solvable models for many-body systems far from equilibrium}\ }(\bibinfo
  {publisher} {Elsevier},\ \bibinfo {year} {2001})\ pp.\ \bibinfo {pages}
  {1--251}\BibitemShut {NoStop}%
\bibitem [{\citenamefont {Yao}\ and\ \citenamefont {Wang}(2018)}]{yao2018edge}%
  \BibitemOpen
  \bibfield  {author} {\bibinfo {author} {\bibfnamefont {S.}~\bibnamefont
  {Yao}}\ and\ \bibinfo {author} {\bibfnamefont {Z.}~\bibnamefont {Wang}},\
  }\bibfield  {title} {\bibinfo {title} {Edge states and topological invariants
  of non-hermitian systems},\ }\href
  {https://doi.org/10.1103/PhysRevLett.121.086803} {\bibfield  {journal}
  {\bibinfo  {journal} {Phys. Rev. Lett.}\ }\textbf {\bibinfo {volume} {121}},\
  \bibinfo {pages} {086803} (\bibinfo {year} {2018})}\BibitemShut {NoStop}%
\bibitem [{\citenamefont {Yokomizo}\ and\ \citenamefont
  {Murakami}(2020)}]{yokomizo2020non}%
  \BibitemOpen
  \bibfield  {author} {\bibinfo {author} {\bibfnamefont {K.}~\bibnamefont
  {Yokomizo}}\ and\ \bibinfo {author} {\bibfnamefont {S.}~\bibnamefont
  {Murakami}},\ }\bibfield  {title} {\bibinfo {title} {Non-bloch band theory
  and bulk--edge correspondence in non-hermitian systems},\ }\href
  {https://doi.org/10.1093/ptep/ptaa140} {\bibfield  {journal} {\bibinfo
  {journal} {Prog. Theor. Exp. Phys.}\ }\textbf {\bibinfo {volume} {2020}},\
  \bibinfo {pages} {12A102} (\bibinfo {year} {2020})}\BibitemShut {NoStop}%
\bibitem [{\citenamefont {Hatano}\ and\ \citenamefont
  {Nelson}(1996)}]{Hatano1996}%
  \BibitemOpen
  \bibfield  {author} {\bibinfo {author} {\bibfnamefont {N.}~\bibnamefont
  {Hatano}}\ and\ \bibinfo {author} {\bibfnamefont {D.~R.}\ \bibnamefont
  {Nelson}},\ }\bibfield  {title} {\bibinfo {title} {Localization transitions
  in non-hermitian quantum mechanics},\ }\href
  {https://doi.org/10.1103/PhysRevLett.77.570} {\bibfield  {journal} {\bibinfo
  {journal} {Phys. Rev. Lett.}\ }\textbf {\bibinfo {volume} {77}},\ \bibinfo
  {pages} {570} (\bibinfo {year} {1996})}\BibitemShut {NoStop}%
\bibitem [{\citenamefont {Kawabata}\ \emph {et~al.}(2020)\citenamefont
  {Kawabata}, \citenamefont {Okuma},\ and\ \citenamefont
  {Sato}}]{Kawabata2020Symplectic_nonBloch}%
  \BibitemOpen
  \bibfield  {author} {\bibinfo {author} {\bibfnamefont {K.}~\bibnamefont
  {Kawabata}}, \bibinfo {author} {\bibfnamefont {N.}~\bibnamefont {Okuma}},\
  and\ \bibinfo {author} {\bibfnamefont {M.}~\bibnamefont {Sato}},\ }\bibfield
  {title} {\bibinfo {title} {Non-bloch band theory of non-hermitian
  hamiltonians in the symplectic class},\ }\href
  {https://doi.org/10.1103/PhysRevB.101.195147} {\bibfield  {journal} {\bibinfo
   {journal} {Phys. Rev. B}\ }\textbf {\bibinfo {volume} {101}},\ \bibinfo
  {pages} {195147} (\bibinfo {year} {2020})}\BibitemShut {NoStop}%
\bibitem [{\citenamefont {{Li}}\ \emph {et~al.}(2020)\citenamefont {{Li}},
  \citenamefont {{Lee}}, \citenamefont {{Mu}},\ and\ \citenamefont
  {{Gong}}}]{Li2020critical}%
  \BibitemOpen
  \bibfield  {author} {\bibinfo {author} {\bibfnamefont {L.}~\bibnamefont
  {{Li}}}, \bibinfo {author} {\bibfnamefont {C.~H.}\ \bibnamefont {{Lee}}},
  \bibinfo {author} {\bibfnamefont {S.}~\bibnamefont {{Mu}}},\ and\ \bibinfo
  {author} {\bibfnamefont {J.}~\bibnamefont {{Gong}}},\ }\bibfield  {title}
  {\bibinfo {title} {{Critical non-Hermitian skin effect}},\ }\href
  {https://doi.org/10.1038/s41467-020-18917-4} {\bibfield  {journal} {\bibinfo
  {journal} {Nat.Commun.}\ }\textbf {\bibinfo {volume} {11}},\ \bibinfo {eid}
  {5491} (\bibinfo {year} {2020})}\BibitemShut {NoStop}%
\bibitem [{\citenamefont {Li}\ \emph {et~al.}(2022)\citenamefont {Li},
  \citenamefont {Cao}, \citenamefont {Chen},\ and\ \citenamefont
  {Yang}}]{Li2023Universal}%
  \BibitemOpen
  \bibfield  {author} {\bibinfo {author} {\bibfnamefont {Y.}~\bibnamefont
  {Li}}, \bibinfo {author} {\bibfnamefont {Y.}~\bibnamefont {Cao}}, \bibinfo
  {author} {\bibfnamefont {Y.}~\bibnamefont {Chen}},\ and\ \bibinfo {author}
  {\bibfnamefont {X.}~\bibnamefont {Yang}},\ }\bibfield  {title} {\bibinfo
  {title} {Universal characteristics of one-dimensional non-hermitian
  superconductors},\ }\href {https://doi.org/10.1088/1361-648X/aca4b4}
  {\bibfield  {journal} {\bibinfo  {journal} {J. Phys.: Condens. Matter}\
  }\textbf {\bibinfo {volume} {35}},\ \bibinfo {pages} {055401} (\bibinfo
  {year} {2022})}\BibitemShut {NoStop}%
\bibitem [{\citenamefont {Kitaev}(2001)}]{Kitaev2001Unpaired}%
  \BibitemOpen
  \bibfield  {author} {\bibinfo {author} {\bibfnamefont {A.~Y.}\ \bibnamefont
  {Kitaev}},\ }\bibfield  {title} {\bibinfo {title} {Unpaired majorana fermions
  in quantum wires},\ }\href@noop {} {\bibfield  {journal} {\bibinfo  {journal}
  {Physics-Uspekhi}\ }\textbf {\bibinfo {volume} {44}},\ \bibinfo {pages} {131}
  (\bibinfo {year} {2001})}\BibitemShut {NoStop}%
\bibitem [{\citenamefont {Katsura}\ \emph {et~al.}(2015)\citenamefont
  {Katsura}, \citenamefont {Schuricht},\ and\ \citenamefont
  {Takahashi}}]{PhysRevB.92.115137}%
  \BibitemOpen
  \bibfield  {author} {\bibinfo {author} {\bibfnamefont {H.}~\bibnamefont
  {Katsura}}, \bibinfo {author} {\bibfnamefont {D.}~\bibnamefont {Schuricht}},\
  and\ \bibinfo {author} {\bibfnamefont {M.}~\bibnamefont {Takahashi}},\
  }\bibfield  {title} {\bibinfo {title} {Exact ground states and topological
  order in interacting kitaev/majorana chains},\ }\href
  {https://doi.org/10.1103/PhysRevB.92.115137} {\bibfield  {journal} {\bibinfo
  {journal} {Phys. Rev. B}\ }\textbf {\bibinfo {volume} {92}},\ \bibinfo
  {pages} {115137} (\bibinfo {year} {2015})}\BibitemShut {NoStop}%
\end{thebibliography}%

\end{document}


\title{Supplemental Material for ``Non-Bloch dynamics and topology in a classical non-equilibrium process"}

\author{Bo Li}
\thanks{These authors contributed equally to this work.}
 \affiliation{ Institute for
Advanced Study, Tsinghua University, Beijing,  100084, China }
 \author{He-Ran Wang}
 \thanks{These authors contributed equally to this work.}
 \affiliation{ Institute for
Advanced Study, Tsinghua University, Beijing,  100084, China }
\author{Fei Song}
\affiliation{ Institute for
Advanced Study, Tsinghua University, Beijing,  100084, China }
\author{Zhong Wang} \altaffiliation{ wangzhongemail@tsinghua.edu.cn }
\affiliation{ Institute for
Advanced Study, Tsinghua University, Beijing,  100084, China }

\maketitle
\onecolumngrid

%
%
%
%
%
%
%
%
%
\hypersetup{linkcolor=black}
\tableofcontents

\section{Free-fermion systems}

In the main text, we impose a constraint on the parameters to eliminate the fermion-density interaction terms in the fermionic representation, allowing for exact solutions.
Here, we emphasize that we adopted this model only for the sake of concreteness: There exists a broad range of stochastic processes supporting free-fermion representations, where the non-Bloch band theory is applicable as well, accompanied by the presence of Majorana zero modes.
Ref.~\cite{Schutz2001Exactly} outlined a method for identifying such models dubbed as ``free-fermion systems'', relying on applying certain similarity transformations on interacting models to obtain free-fermion Hamiltonians.
In the following, we provide a brief sketch of this method.\\
\indent We focus on stochastic processes of single-specie particles with local processes involving two adjacent sites. In general, the Hamiltonian can be written as
\begin{eqnarray}
\mathcal H=\sum_j h_{j,j+1},
\end{eqnarray}
where $h_{j,j+1}$ is a $4\times 4$ matrix that collects interactions in terms of $\sigma_j^\alpha\sigma_{j+1}^\beta$, $\alpha,\beta\in\{0,x,y,z\}$. We aim to search for $\mathcal H$ that can be connected to a free-fermion Hamiltonian $\mathcal H^{ff}$ through a similarity transformation $\mathcal B$, i.e.,
\begin{eqnarray}
    \mathcal H=\mathcal B\mathcal H^{ff}\mathcal B^{-1}.
\end{eqnarray}
Here, we concentrate on local transformations of the homogeneous form $\mathcal B=B^{\bigotimes L}$, where $B$ is a $2\times2$ matrix, and $L$ is the number of sites. The strategy starts from the most general $\mathcal{H}^{ff}$ and $\mathcal B$ to find $\mathcal H$ by imposing the probability conservation condition. To this end, we consider $\mathcal H^{ff}=-\sum_j h^{ff}_{j,j+1}$ with
\begin{eqnarray}
 h^{ff}_{j,j+1}=c+D_1 \sigma_j^+\sigma_{j+1}^-  +D_2 \sigma_j^-\sigma_{j+1}^+ +\mu_1\sigma_j^+\sigma_{j+1}^++\mu_2\sigma_j^-\sigma_{j+1}^-+h_1n_j+h_2 n_{j+1}.
\end{eqnarray}
The probability conservation reads $\langle I|\mathcal H=0$, which indicates $\langle I|\mathcal{B} h^{ff}_{j,j+1}=0$. \\
\indent We directly list all the solutions, which are classified into six categories as follows \cite{Schutz2001Exactly}:
\begin{eqnarray*}
 h^{I,A}_{j,j+1}=\left(
 \begin{array}{cccc}
\bullet& 0& 0& (1-\nu)(D_R+D_L)  \\
0&\bullet& D_R& 0\\
0& D_L& \bullet& 0\\
\nu(D_L+D_R)& 0& 0& \bullet
 \end{array}
 \right),
\qquad
 h^{I,B}_{j,j+1}=\left(
 \begin{array}{cccc}
\bullet& 0& 0& (1-\nu)(D_R+D_L)  \\
0&\bullet& D_R& \nu D_R\\
0& D_L& \bullet& \nu D_L\\
0 & 0& 0& \bullet
 \end{array}
 \right),
\end{eqnarray*} 
\begin{eqnarray*}
 h^{I,C}_{j,j+1}=\left(
 \begin{array}{cccc}
\bullet& 0& 0& 0  \\
0&\bullet& D_R& D_R\\
0& D_L& \bullet& D_L\\
0 & \nu D_L& \nu D_R& \bullet
 \end{array}
 \right),
 h^{I,D}_{j,j+1}=\left(
 \begin{array}{cccc}
\bullet& a\nu& a\nu^{-1}& 0  \\
0&\bullet& 0& 0\\
0& 0& \bullet& 0\\
0 & b\nu^{-1} & b \nu & \bullet
 \end{array}
 \right),
 h^{II,A}_{j,j+1}=\left(
 \begin{array}{cccc}
\bullet& a& b& 0  \\
c&\bullet& 0& d\\
d& 0& \bullet& c\\
0 & b & a & \bullet
 \end{array}
 \right),
 h^{II,B}_{j,j+1}=\left(
 \begin{array}{cccc}
\bullet& a& a& 0  \\
b&\bullet& 0& a\\
b& 0& \bullet& a\\
0 & b & b & \bullet
 \end{array}
 \right).
\end{eqnarray*}
We follow the notations in Ref. \cite{Schutz2001Exactly} to label and parameterize the models. All the off-diagonal matrix elements are demanded to be non-negative. We abbreviate the diagonal elements as $\bullet$, since they can be automatically filled in according to the probability conservation. Notably, the model we studied in the main text exactly corresponds to $h^{I,A}_{j,j+1}$.\\
\indent Now we discuss the physical implications of the above solved free-fermion systems. To this end, we summarize the correspondence between off-diagonal matrix elements $w_{ij}$ (with $i$ denoting the column index and $j$ the row), the microscopic processes, and the Pauli-matrix representations in Table.~\ref{tab:reaction_diffusion}. 
Exact solutions for the free-fermion systems are not attainable \textit{in prior}, since some of them contain non-zero $\sigma^z_j\sigma^z_{j+1}$ terms which map to fermion-density interactions through Jordan-Wigner transformation (i.e., $h_{j,j+1}^{I,C}, h_{j,j+1}^{I,D}, h_{j,j+1}^{II,A}$ and $h_{j,j+1}^{II,B}$). In addition, the presence of non-zero $\sigma^\pm_j n_{j+1}$ terms in some models explicitly breaks parity symmetry, which is not commonly witnessed in free-fermion models. Despite all these unconventional features, the free-fermion systems can be mapped to quadratic Hamiltonians through certain similarity transformations. Their free-fermion counterparts typically involve non-reciprocal hoppings and/or pairings, necessitating the applications of the non-Bloch band theory to solve the OBC spectrum. Furthermore, since parity symmetry is restored in quadratic Hamiltonians, the existence of Majorana zero modes and the dynamical crossover in asymptotic dynamics should be expected as well. This highlights the general applications of the non-Bloch band theory in characterizing anomalous dynamics in a wide spectrum of stochastic processes.\\

\begin{table}[ht]
    \centering
\begin{tabular}{c|c|c}
\hline
Process & Rate &Transition matrix  \\
\hline
Diffusion $l\rightarrow k$\qquad $\emptyset$A$\rightarrow$ A$\emptyset$& $w_{32}$ &$\sigma^-_k\sigma_l^+$ \\
Diffusion $k\rightarrow l$\qquad $A\emptyset$$\rightarrow$ $\emptyset$A& $w_{23}$ &$\sigma^+_k\sigma_l^-$ \\
Pair annihilation\qquad $A A\rightarrow\emptyset\emptyset$& $w_{14}$ &$\sigma^+_k\sigma_l^+$ \\
Pair creation\qquad $\emptyset\emptyset\rightarrow AA$& $w_{41}$ &$\sigma^-_k\sigma_l^-$ \\
Fusion on $k$\qquad
 $AA\rightarrow A\emptyset$& $w_{34}$ & $n_k\sigma_l^+$\\
Fusion on $l$\qquad
 $AA\rightarrow \emptyset A$& $w_{24}$ & $\sigma_k^+ n_l$\\
Branching on $k$\qquad
 $\emptyset A\rightarrow A A$& $w_{42}$ & $\sigma_k^- n_l$\\
Branching on $l$\qquad
 $A\emptyset \rightarrow A A$& $w_{43}$ & $n_k\sigma_l^-$\\
 Death on $k$ \qquad
$A\emptyset\rightarrow\emptyset\emptyset$ & $w_{13}$ & $\sigma_k^+(1-n_l)$ \\
 Death on $l$ \qquad
$\emptyset A\rightarrow\emptyset\emptyset$ & $w_{12}$ & $(1-n_k)\sigma_l^+$ \\
Birth on $k$ \qquad
$\emptyset\emptyset\rightarrow A\emptyset$ & $w_{31}$ & $\sigma_k^-(1-n_l)$ \\
Birth on $l$ \qquad
$\emptyset\emptyset\rightarrow\emptyset A$ & $w_{21}$ & $(1-n_k)\sigma_l^-$ \\
\hline
\end{tabular}
    \caption{Two-site reaction-diffusion process on a pair of site $(k,l)$ are given by quantum-spin representation with corresponding rates. $A$ and $\emptyset$ stand for the occupied and empty site, respectively.}
\label{tab:reaction_diffusion}
\end{table}

\section{Diagonalization of the Hamiltonian}
In this section, we elaborate the diagonalization process for the BdG Hamiltonian and make connections with the Majorana representation, where the Majorana-like zero modes are analyzed in details. 

\subsection{Particle-hole symmetry and diagonalization}

The BdG Hamiltonian matrix naturally inherits a particle-hole symmetry (PHS) from its basis, which can be clearly seen by applying the symmetry transformation, $\Psi^\dagger_i=(\Sigma_x)_{ij}\Psi_j$ with $\Sigma_x=\sigma_x\bigotimes\mathbbm{1}_L$, to the Hamiltonian,
\begin{eqnarray}
\mathcal H=\sum_{i,j}\Psi^\dagger_i H_{ij}\Psi_j=\sum_{i,j}\sum_{k,l}(\Sigma_x)_{il}\Psi_l H_{ij}(\Sigma_x)_{kj}\Psi_k^\dagger=\sum_{k,l}\delta_{lk}(\Sigma_x H^T\Sigma_x)_{kl}-\Psi^\dagger(\Sigma_x H^T\Sigma_x)\Psi,
\end{eqnarray}
where the first term in the last step vanishes due to  $\text{Tr}[\Sigma_x H^T\Sigma_x]=\text{Tr}[H]=0$. Therefore, the particle-hole symmetry requires 
\begin{eqnarray}\label{eq:PHsymmetry}
\Sigma_xH^T\Sigma_x=-H.
\end{eqnarray}
In general, the Hamiltonian can be diagonalized by a transformation matrix $T$
\begin{eqnarray}\label{eq:diagonalization}
T H T^{-1}=E=\text{Diag}\{\varepsilon_1,\varepsilon_2,\cdots,\varepsilon_{2L}\}.
\end{eqnarray}
The diagonalization in Eq.~\eqref{eq:diagonalization} is equivalent to bi-orthogonal eigen-equations of the matrix $H$ 
\begin{subequations}
\begin{align}
 H|\psi_n\rangle=\varepsilon_n|\psi_n\rangle\label{eq:righteigenstate},\\
\langle\bar{\psi}_n|H=\langle\bar{\psi}_n|\varepsilon_n, \label{eq:lefteigenstate} 
\end{align}
\end{subequations}
where $\langle j|\psi_{n}\rangle=(T^{-1})_{jn}$, $\langle\bar{\psi}_{n}|j\rangle=T_{nj}$. Here, the left and right eigenvectors are not Hermitian conjugate to each other due to the non-Hermiticity. However, they form a pair of biorthogonal basis, i.e., $\langle\bar{\psi}_m|\psi_n\rangle=\delta_{mn}$.   
By transposing Eq.~\eqref{eq:lefteigenstate} and using the PHS $CH^TC^{-1}=-H$ with $C=\Sigma_x$, the eigen-equation is transformed to
\begin{eqnarray}
H (C |\bar{\psi}_n\rangle^*)=-\varepsilon_n (C |\bar{\psi}_n\rangle^*).
\end{eqnarray}
Therefore, PHS puts eigenstates into pairs with opposite energies.
Furthermore, elements of the Hamiltonian matrix are purely real numbers, $H^\ast=H$, which further induces the following relations
\begin{equation}
H|\psi_n\rangle^*= \varepsilon^\ast_n|\psi_n\rangle^*,\qquad H(C|\bar{\psi}\rangle)=-\varepsilon^*_n(C|\bar{\psi}\rangle).
\end{equation}
We summarize the connections between eigenvalues and eigenstates as follows 
\begin{eqnarray}
(\varepsilon_n,\varepsilon_n^\ast,-\varepsilon_n,-\varepsilon_n^\ast)\Rightarrow
(|\psi_n\rangle, |\psi_n\rangle^*, C|\bar{\psi}_n\rangle^*, C|\bar{\psi}_n\rangle).
\end{eqnarray}

With a proper arrangement, the energy matrix in Eq.~\eqref{eq:diagonalization} can be explicitly expressed as 
\begin{eqnarray}\label{eq:eigenenergy}
E=\text{Diag}\{\varepsilon_1,\cdots, \varepsilon_L,-\varepsilon_1,\cdots,-\varepsilon_L\}.
\end{eqnarray}
The eigenstate with the correct order following from the energy matrix should sustain the relation up to a phase factor: $|\psi_{L+n}\rangle=C |\bar{\psi}_n\rangle^*$ and $|\psi_n\rangle=|\bar{\psi}_{L+n}\rangle^*$ for $n<L$.


After diagonalization, the Hamiltonian can be written on the eigenstates ($\xi_R=T\Psi$ and $\xi_L^+=\Psi^\dagger T^{-1}$) as
\begin{eqnarray}
\mathcal H=\frac{1}{2}\sum_{n=1}^{2L}E_{n}\xi_{L,n}^+\xi_{R,n}.
\end{eqnarray}
It is readily to verify that the new operators still respects the anticommutation relations $\{\xi_{L,m}^+,\xi_{R,n}\}=\delta_{mn}$, though $\xi_L^+$ and $\xi_R$ are not Hermitian conjugate pair. In addition, from $\{\Psi_i,\Psi_j\}=(\Sigma_x)_{ij}$, it is straightforward to obtain
\begin{eqnarray}
\{\xi_{R,m},\xi_{R,n}\}=(T\Sigma_x T^t)_{mn}=(\Sigma_x)_{mn}.
\end{eqnarray}
This relation combined with the eigenvalue matrix Eq.~\eqref{eq:eigenenergy} suggest the particle-hole nature of the new basis 
\begin{eqnarray}\label{eq:Eigenbasis}
&&\xi_{R}=(\alpha_1,\cdots,\alpha_L,\bar{\alpha}_1^\dagger,\cdots,\bar{\alpha}_L^\dagger)^T,\nonumber\\
&&\xi_L^+=(\bar{\alpha}_1^\dagger,\cdots,\bar{\alpha}_L^\dagger,\alpha_1,\cdots,\alpha_L).
\end{eqnarray}
Here, $\bar{\alpha}_i^\dagger\neq \alpha_i^\dagger$, $\{\bar{\alpha}^\dagger_i,\alpha_j\}=\delta_{ij}$, and $\{\alpha_i,\alpha_j\}=\{\bar{\alpha}^\dagger_i,\bar{\alpha}^\dagger_j\}=0$. 
Therefore, the Hamiltonian can be represented by the quasi-particle operators
\begin{eqnarray}
\mathcal H=\sum_{n=1}^L\varepsilon_n\bar{\alpha}^\dagger_n\alpha_n.
\end{eqnarray}

\subsection{Particle-hole symmetry and zero modes in the Majorana presentation}

We rewrite the Fermion creation and annihilation operators under the Majorana representation
\begin{eqnarray}\label{eq:Majorana_transformation}
\left(
\begin{array}{cc}
    c_i\\ c_i^\dagger
\end{array}\right)=U
\left(
\begin{array}{cc}
    a_i\\ b_i
\end{array}\right)\qquad \text{with}\qquad
U=\frac{1}{\sqrt{2}}\left(
\begin{array}{cc}
     1&i  \\
     1&-i 
\end{array}\right),
\end{eqnarray}
where $a_i=a_i^\dagger$, $b_i=b_i^\dagger$, and $a_i^2=b_i^2=1$. As a result, the Nambu basis is transformed as $\Psi=V\Gamma$ where $V=U\bigotimes\mathbbm{1}_L$ satisfies $VV^\dagger=1$, and $\Gamma=\{a_1,a_2,\cdots,a_L, b_1,b_2,\cdots b_L\}$ with $\{\Gamma_i,\Gamma_j\}=\delta_{ij}$. In the Majorana basis, the Hamiltonian matrix becomes
\begin{eqnarray}\label{eq:MajoranaHamiltonian}
H_M=V^\dagger H V.
\end{eqnarray}
The particle-hole symmetry Eq.~\eqref{eq:PHsymmetry} now manifests as $H_M^T=-H_M$.
On the Majorana basis, if  $|u_n\rangle$ and  $\langle \bar{u}_n|$ form a pair of right and left eigenstates, i.e., $H_M|u_n\rangle=\varepsilon_n|u_n\rangle$ and  $\langle \bar{u}_n|H_M=\langle\bar{u}_n|\varepsilon_n$, the eigenstate with opposite energy can be found as
\begin{eqnarray}
H_M|\bar{u}_n\rangle^\ast=-\varepsilon_n|\bar{u}_n\rangle^\ast.
\end{eqnarray}

A special case arises when $\varepsilon_1=0$. Because nonzero energies in particle-hole basis appear in pairs with opposite sign, while the zero energy is paired with itself, which leads to the degeneracy at zero energy, i.e., there are two Majorana-like zero modes forming one (non-local) fermion. Within the zero-energy subspace, the particle-hole symmetry can be implemented in a different way due to the degeneracy. Given $H_M|u^{(0)}_1\rangle=0$, the PHS leads $\langle u^{(0)}_1|^\ast H_M=0$, so that $\langle\bar{u}_1^{(0)}|=\langle u^{(0)}_1|^\ast$. 
Moreover, the eigenstate can be always normalized as $\sum_i \langle i|u_1^{(0)}\rangle^2=1$. The other left and right pair $(\langle \bar{u}_2^{(0)}|, |u_2^{(0)}\rangle)$ corresponding to zero energy satisfy the same relation. These state can always be chosen to be orthogonal  $\langle\bar{u}_1^{(0)}|u_2^{(0)}\rangle=0$; otherwise, if $\langle\bar{u}_1^{(0)}|u_2^{(0)}\rangle=\kappa$, we can replace the second zero state can be replaced by $|u^{(0)}_2\rangle-\kappa|u^{(0)}_1\rangle$ make them orthogonal to each other.

From this perspective, the two zero modes are identified as Majorana fermions due to the following relations
\begin{eqnarray}
&&\gamma_1=\bra{i}u_{1}^{(0)}\rangle \Gamma_i=\bra{i}\bar{u}_{1}^{(0)}\rangle\Gamma_i=\bar{\gamma}_1^\dagger,\nonumber\\
&&\gamma_2=(\bra{i}u_{2}^{(0)}-\kappa \bra{i}u_{1}^{(0)}) \Gamma_i=(\bar{u}_{2}^{(0)}-\kappa\bar{u}_{1}^{(0)})\Gamma_i=\bar{\gamma}_2^\dagger.
\end{eqnarray}
It is straightforward to check that
\begin{eqnarray}
\gamma_1^2=\frac{1}{2}\{\gamma_1,\gamma_1\}=\frac{1}{2}\sum_{i,j}\bra{i}\bar{u}_{1}^{(0)}\rangle\bra{j}\bar{u}_{1}^{(0)}\rangle\{\Gamma_i,\Gamma_j\}=\frac{1}{2}\sum_i\bra{i}\bar{u}_{1}^{(0)}\rangle^2=\frac{1}{2},
\end{eqnarray}
which also works for $\gamma_2$. These two zero modes can be utilized to construct a zero-energy fermion, i.e., $\alpha_1=(\gamma_1+i\gamma_2)/\sqrt{2}$ and $\bar{\alpha}_1^\dagger=(\gamma_1-i\gamma_2)/\sqrt{2}$.
It is worth noting that the Majorana-like zero modes found above are not unique, that they can be redefined by the following transformation
\begin{eqnarray}
\left(\begin{array}{cc}
|u_1^\prime\rangle \\
|u_2^\prime\rangle
\end{array}\right)=
\left(\begin{array}{cc}
\cos\theta &-\sin\theta\\
\sin\theta&\cos\theta
\end{array}\right)
\left(\begin{array}{cc}
|u_1\rangle \\
|u_2\rangle
\end{array}\right).
\end{eqnarray}
However, this transformation can only contribute to an extra phase factor to the zero-energy fermion, i.e., $\alpha_1\rightarrow e^{i\theta}\alpha_1$ and $\bar{\alpha}^\dagger_1\rightarrow e^{-i\theta}\bar{\alpha}^\dagger_1$. 

\section{Spectrum}\label{sec:continuum bands}
In this section, we show that the continuum band under OBC can be analytically calculated by the non-Bloch band theory \cite{yao2018edge, yokomizo2020non}. When only continuum bands are concerned, the model is shown to be equivalent to a Hatano-Nelson (HN) model \cite{Hatano1996} without disorder. 
To this end, we first decompose each fermion into a pair of Majorana fermions following Eq.~\eqref{eq:Majorana_transformation},
then the original Hamiltonian is represented as
\begin{eqnarray}\label{eq:Majorana_form}
\mathcal H=\sum_j-[(\delta_2-\delta_1)a^\prime_{j+1}a^\prime_j+(\delta_1+\delta_2)b^\prime_{j+1}b^\prime_j+i2sa^\prime_{j+1}b^\prime_j]-\delta_1(1+2ia^\prime_jb^\prime_j),
\end{eqnarray}
where boundary terms are neglected. 
Now, we recombine the Majorana operators into a set of new fermions 
\begin{eqnarray}\label{seq:f-fermion}
f_j=(b^\prime_{j-1}+ia^\prime_j)/\sqrt{2}, \qquad f_j^\dagger=(b^\prime_{j-1}-ia^\prime_j)/\sqrt{2},
\end{eqnarray}
through which Eq.~\eqref{eq:Majorana_form} is reduced to $\mathcal H=\mathcal H_{\text{HN}}-2\delta_1\sum_j f_{j+1}^\dagger f_j^\dagger+\text{const}.$, with
\begin{eqnarray}
\mathcal H_{\text{HN}}=\sum_j (2s f_j^\dagger f_j+(\delta_1-\delta_2)f^\dagger_{j+1}f_j+(\delta_1+\delta_2)f_j^\dagger f_{j+1}).
\end{eqnarray}
With the absence of pair annihilation terms $f_{j+1}f_j$, the BdG form of $\mathcal H$ in the $(f_j,f_j^\dagger)$ basis is reduced to an upper triangle matrix, and its continuum spectrum is exactly captured by the diagonal part $\mathcal H_{\text{NH}}$. Although we neglected the boundary terms here, it is actually important for sustaining such an upper-triangular form, which will be discussed in detail in the next section. The above derivation is performed for an infinite system, the conclusion would apply to the continuum bands for both periodic and open boundary conditions.

Under PBC, the band is directly given by the Bloch Hamiltonian
\begin{eqnarray}
E_{\text{PBC}}(k)=H_{\text{HN}}(k)=2s+(\delta_1-\delta_2)e^{-ik}+(\delta_1+\delta_2)e^{ik},
\end{eqnarray}
where the momentum $k$ takes different discrete values for even or odd sector, depending on either antiperiodic or periodic boundary condition is taken.

However, the presence of NHSE under OBC necessitates the non-Bloch band theory to solve the spectrum \cite{yao2018edge, yokomizo2020non}. Both of the continuum band and GBZ can be extracted from the characteristic equation
\begin{eqnarray}
\text{det}[H_{HN}(\beta)-E]=0
\end{eqnarray}
where $H_{HN}(\beta)$ is obtained by substituting $e^{ik}\rightarrow\beta$ into  $H_{NH}(HN)$. This equation yields eigenvalues:
\begin{eqnarray}\label{eq:OBCband_beta}
E_{\text{OBC}}(\beta)=2s+(\delta_1-\delta_2)\beta^{-1}+(\delta_1+\delta_2)\beta,
\end{eqnarray}
where the corresponding GBZ is a circle with radius  $\sqrt{|\delta_1-\delta_2|/|\delta_1+\delta_2|}$. Therefore, the points on the GBZ can be parameterized by 
\begin{eqnarray}\label{eq:beta}
\beta=\sqrt{|\delta_1-\delta_2|/|\delta_1+\delta_2|}e^{i\phi}.
\end{eqnarray}
Plugging this expression into Eq.~\eqref{eq:OBCband_beta} we obtain the OBC continuum spectrum
\begin{eqnarray}\label{eq:MZM_localization}
E_{\text{OBC}}(\phi)=2s+\sqrt{|\delta_1^2-\delta_2^2|}[\text{sgn}(\delta_1-\delta_2)e^{-i\phi}+\text{sgn}(\delta_1+\delta_2)e^{i\phi}].
\end{eqnarray}

Additionally, the wavefunction of the zero mode should also be governed by the same equation in the bulk. We can extract the localization length of isolated zero modes from Eq.~\eqref{eq:OBCband_beta} by setting $E(\beta)=0$, resulting in
\begin{eqnarray}
\beta_\pm=\frac{-s\pm \sqrt{s^2-(\delta_1^2-\delta_2^2)}}{\delta_1+\delta_2}.
\end{eqnarray}
The zero modes localize at left (or right) boundary with localization length $|\ln|{\beta_l}||^{-1}$ if $|\beta_l|<1$ (or $>1$) with $l=\pm$. This calculation is in analogous to the differential equation determining the topological edge (surface) mode in Hermitian cases. Moreover, it is worth noting a special case : either $|\beta_+|$ or $|\beta_-|$ will take a unit value if $s^2=\delta_1^2$, corresponding to a defective zero mode.

\section{The zero mode and topology}
This section will be devoted to the band topology, especially the stability of the zero mode and the subtlety of topological phase transition. To uncover the feature of the zero mode, we first write down the full Hamiltonian, including boundary terms, in the  $f$-fermion representation [Eq.~\eqref{seq:f-fermion}]. To this end, we first clearly recap the Hamiltonian under OBC
\begin{eqnarray}\label{seq:OBC_Hamiltonian}
\mathcal H=-\sum_{j=1}^{L-1}\epsilon c_j^\dagger c^\dagger_{j+1}+\epsilon^\prime c_{j+1}c_{j}+h c^\dagger_{j+1}c_j+h^\prime c^\dagger_j c_{j+1}+(\epsilon^\prime-\epsilon)\sum_{j=2}^{L-1}c^\dagger_j c_j+(h-\epsilon)c_1^\dagger c_1+(h^\prime-\epsilon)c_L^\dagger c_L+\text{const.}.
\end{eqnarray}
Notice that the on-site potential for the first and last site are different from others in the bulk.  The transformation Eq.~\eqref{eq:Majorana_transformation} and~\eqref{seq:f-fermion} can be combined into a compact form as: 
\begin{eqnarray}\label{seq:f-transformation}
 \left(\begin{array}{cc}
c_j\\ c^\dagger_j
 \end{array}\right)=
 W
\left(
\begin{array}{cc}
    f_j\\ f_j^\dagger
\end{array}\right)+W^\dagger
\left(
\begin{array}{cc}
    f_{j+1}\\ f^\dagger_{j+1}
\end{array}\right)\qquad\text{with}\qquad
W=\frac{i}{2}\left(
\begin{array}{cc}
    -1 &1  \\
    -1 &1 
\end{array}
\right),
\end{eqnarray}
where we identify $f_{L+1}=f_1$ which is composed of non-local Majorana fermions $f_1=(b^\prime_L+i a^\prime_1)/\sqrt{2}$. Inserting Eq.~\eqref{seq:f-transformation} into~\eqref{seq:OBC_Hamiltonian}, the full Hamiltonian reads
\begin{eqnarray}
 \mathcal H=&& \sum_{j=2}^{L-1} [2s f_j^\dagger f_j+(\delta_1-\delta_2)f^\dagger_{j+1}f_j+(\delta_1+\delta_2)f_j^\dagger f_{j+1}-2\delta_1f_{j+1}^\dagger f_j^\dagger]+2s f_L^\dagger f_L\nonumber\\
 &&+(\delta_1-\delta_2)(f_2^\dagger f_1+f_1^\dagger f_2^\dagger)
 +(\delta_1+\delta_2)(f_L^\dagger f_1+f_L^\dagger f_1^\dagger)\nonumber\\
 =&&\frac{1}{2}(F^\dagger,F)
 \left(
 \begin{array}{cc}
    \mathbf A  &\mathbf B  \\
    \mathbf{0}  &-\mathbf{A}^T 
 \end{array}
 \right)\left(
 \begin{array}{cc}
    F \\
    F^\dagger 
 \end{array}
 \right),
\end{eqnarray}
where $F=\{f_1,f_2,\cdots,f_L\}$ and
\begin{eqnarray}
\mathbf{A}=\left(
\begin{array}{cccccc}
0&0&0&\cdots&\cdots&\cdots\\ 
t_-&2s&t_+&\cdots&\cdots&\cdots\\
0&t_-&2s&t_+&\cdots&\cdots\\
\vdots&\vdots&\vdots&\ddots&\vdots&\vdots\\
0&0&0&\cdots&2s&t_+ \\
t_+ & 0& \cdots&\cdots&t_-&2s
\end{array}\right),
\end{eqnarray}
with $t_\pm=\delta_1\pm\delta_2$. Due to the upper-triangular matrix form under the $f$-fermion representation, the spectrum is fully determined by $\mathbf{A}$ (and $-\mathbf{A}^T$). Since $\text{Det}(\mathbf A)=0$, a zero mode always exists, regardless of of specific parameter values. 
This implies the existence of an exact zero mode even for the gapless bulk spectrum, i.e., $s^2<\delta_1^2-\delta_2^2$, which contrasts the conventional scenario of topological phases where the zero mode or other topological edge modes lie in the finite-width band gap. 
This anomalous feature is closely related to the decoupled structure of two bands in real space (the absence of lower-triangular matrix elements). By tuning parameters, the bands cross without any repulsion, preventing a topological transition. Thus, it would be questionable to relate the topological phase transition with the continuous-spectrum gap closing at $s^2=\delta_1^2-\delta_2^2$.\\
\indent To address the problem, we explore parameters beyond the constrained subspace studied in the main text. Specifically, we consider changing the on-site potential on the first and last site to be the same as the bulk values [i.e., $(h-\epsilon)c^\dagger_1 c_1\rightarrow (\epsilon^\prime-\epsilon)c^\dagger_1 c_1$ and $(h^\prime-\epsilon)c^\dagger_L c_L\rightarrow (\epsilon^\prime-\epsilon)c^\dagger_L c_L$], which preserves the particle-hole symmetry but violates the probability conservation. 
We will show that applying such a non-physical perturbation proves beneficial for understanding the topological stability of the zero mode.  
In Fig.~\ref{fig:zero_mode} below, we display the OBC spectrum with and without the boundary perturbations. In the left subfigure,  the unperturbed model always holds a zero mode, even in the ``gapless phase", with the gap closing point at $s^2=\delta_1^2-\delta_2^2$.  However, by adding only the boundary perturbations, the bulk spectrum changes dramatically as shown in the right subfigure.
Now the gap closing point moves to $s=\delta_1$, accompanied by the vanishing/appearance of the zero mode. We emphasize that $s=\delta_1$ is exactly the gap closing point of the original Hamiltonian under PBC, as will be demonstrated later on.
Notice that for $s<\delta_1$, the seemingly gapped region results from the finite-size effect. For $s>\delta_1$, the zero modes split with an exponentially small energy difference, rather than being exact at zero.
The instability of the energy spectrum against a local perturbation implies that the gap closing point of the original Hamiltonian \eqref{seq:OBC_Hamiltonian} fails to signal the topological phase transition, and thus calls for a stable topological invariant to characterize the transition.\\
\indent First, we show that the instability of the spectrum can be demonstrated by the non-Bloch band theory, referred as the critical skin effect \cite{Kawabata2020Symplectic_nonBloch,Li2020critical}. 
The GBZs for the original Hamiltonian are two circles with inverse radius, corresponding to the decoupled particle and hole band with opposite localization directions. 
Small perturbations on the two ends can be transformed to effective couplings between the bands under the $f$-fermion representation, introducing non-zero lower-triangular matrix elements. In the real space, the two independent $f$-fermion chains are coupled by attaching the end points on the same side, leading to an effective PBC.
Consequently, with small and local inter-band couplings, the OBC spectrum undergoes sudden change, approximately following the PBC spectrum, which is stable against local perturbations.
This also elucidates why the phase transition point in Fig.~\ref{fig:zero_mode}(b) coincides with the PBC gap closing point. \\
\indent Next, regarding the topological invariant, it should be associated with the stable gap closing point.
According to the non-Bloch band theory for generic particle-hole symmetric models~\cite{Kawabata2020Symplectic_nonBloch,Li2023Universal}, at the topological phase transition, the particle and hole non-Bloch bands typically cross at the momentum $\beta=\pm 1$. Notice that this momentum also lies on the BZ, implying the simultaneous gap closing of the PBC spectrum. 
Therefore, the topological invariant under PBC can also predict the topological phase transition point, though the zero mode only exists under OBC. 
For our model, the Bloch Hamiltonian is given by $H(k)=i2\delta_2\sin k\sigma_0-i2\delta_1\sin k\sigma_x+2s\sin k\sigma_y-2(\delta_1+s\cos k)\sigma_z$ where $\sigma_j (j=0,x,y,z)$ are Pauli matrices.  The PBC topological invariant~\cite{Kitaev2001Unpaired} is given based on the Majorana-basis matrix  $i B(k)=U^\dagger H(k) U$ with $U$ defined in Eq.~\eqref{eq:Majorana_transformation}:
\begin{eqnarray}
 \nu=\text{sgn}[\text{Pf}(B(0))\text{Pf}(B(\pi))]=\text{sgn}(\delta^2_1-s^2).   
\end{eqnarray}
The index $\nu=\pm 1$ respectively signals the topological trivial and non-trivial phases. Indeed, when $s^2>\delta_1^2$, the system is in the non-trivial phase that maintains a robust zero mode, aligning with Fig.~\ref{fig:zero_mode}.

\begin{figure}
\centering
\includegraphics[width=0.8\linewidth]{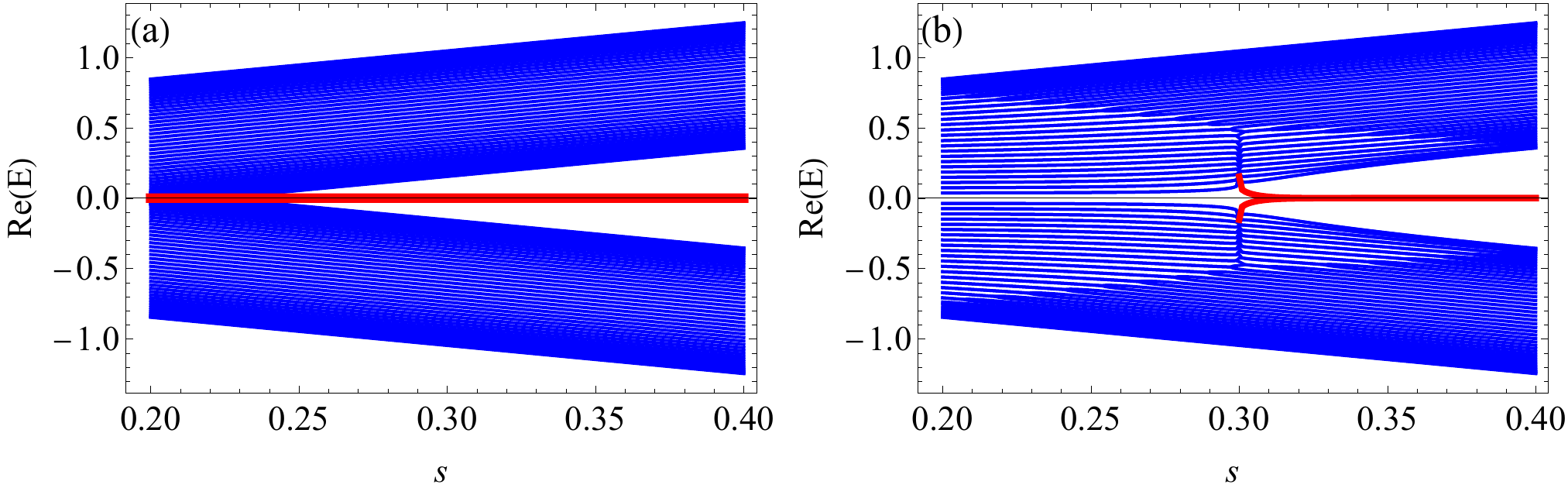}
\caption{ (a) Real part of spectrum as a function of $s$ for the original Hamiltonian Eq.~\eqref{seq:OBC_Hamiltonian}. (b) Real part of spectrum for the Hamiltonian with modified boundary terms, i.e., unifying the on-site potential for all sites. In the plots, $\delta_1=0.3$, $\delta_2=0.2$, and $L=50$.}
    \label{fig:zero_mode}
\end{figure}

\section{Steady state}

In this section we will gain some understanding of the steady sate. The probability conservation requires that
\begin{eqnarray}
\langle I|\mathcal H=0
\end{eqnarray}
where $\langle I|=\sum_{\eta }\langle \eta|$, $\langle\eta |=\langle n_1,n_2,\cdots,n_L|$ with $n_i=0$ or $1$ labeling the empty or occupied state for a given site. Since the Hamiltonian preserves the parity of particle number, this relation should be satisfied individually in even and odd-parity space, i.e., the left steady states are given by the projection of $\langle I|$ onto different parity sectors. 
In the following, we will verify this statement from the viewpoint of minimising the energy of the system \cite{PhysRevB.92.115137} (searching the eigenenergy with the smallest real part).

We first consider the OBC case. The Hamiltonian is neighboring sites 
\begin{eqnarray}
\mathcal H=\sum_{j=1}^{L-1}h_{j,j+1}+(L-1)\epsilon
\end{eqnarray}
with
\begin{eqnarray}
h_{j,j+1}= -(\epsilon c_j^\dagger c^\dagger_{j+1}+\epsilon^\prime c_{j+1}c_{j}+h c^\dagger_{j+1}c_j+h^\prime c^\dagger_j c_{j+1})+(h-\epsilon)c_j^\dagger c_j+(h^\prime-\epsilon)c^\dagger_{j+1}c_{j+1}.
\end{eqnarray}
Searching for the left steady state of the system  is equivalent to finding  the right steady state of $\mathcal H^\dagger$. Each local Hamiltonian $h_{j,j+1}$ preserves the parity, such that under even-parity basis  $(|0_j0_{j+1}\rangle,|1_j1_{j+1}\rangle)$ and odd-parity $(|0_j1_{j+1}\rangle,|1_j0_{j+1}\rangle)$, $h_{j,j+1}^\dagger$ is separately written as
\begin{eqnarray}
(h_{j,j+1}^{\dagger})^{\text{even}}=\left(\begin{array}{cc}
0&-\epsilon\\
-\epsilon^\prime&\epsilon^\prime-\epsilon
\end{array}\right),\qquad
(h_{j,j+1}^\dagger)^{\text{odd}}=\left(\begin{array}{cc}
h^\prime-\epsilon&-h^\prime\\
-h&h-\epsilon
\end{array}\right).
\end{eqnarray}
Here, $(h_{j,j+1}^\dagger)^{\text{even}}$ is diagonalized by $(-h^\prime/h,1)^t$ and $(1,1)^t$ with eigenvalue $\epsilon^\prime$ and $-\epsilon$, respectively; $(h_{j,j+1}^\dagger)^{\text{odd}}$ is diagonalized by $(-\epsilon^\prime/\epsilon,1)^t$ and $(1,1)^t$ with the corresponding eigenvalue $\epsilon^\prime$ and $-\epsilon$. The lowest-eigenvalue state is degenerate, hence we are able to obtain a linear combination factorized as
\begin{eqnarray}
(1+c_j^\dagger c_{j+1}^\dagger)|\text{vac}\rangle \pm(c^\dagger_j+c^\dagger_{j+1})|\text{vac}\rangle=(1\pm c_j^\dagger)(1\pm c_{j+1}^\dagger)|\text{vac}\rangle.
\end{eqnarray}
As the factorized form decouples different sites, it can simultaneously minimise all dimer components in the Hamiltonian, thus to reach the lowest energy of the full Hamiltonian. Therefore, the left state minimising the energy is constructed as 
\begin{eqnarray}\label{eq:steady_state_construction}
\langle \psi^{\pm}_L|=\langle 0|\prod_{j=1}^{L}(1\pm c_j).
\end{eqnarray}
However, $\langle\psi^\pm_L|$ falls into neither the even nor the odd parity, and we construct the left steady state with even or odd parity as below
\begin{equation}
\langle G^e_L|=\frac{1}{\mathcal N^e}(\langle\psi_L^+|+\langle\psi_L^-|),\qquad
\langle G^o_L|=\frac{1}{\mathcal N^o}(\langle\psi_L^+|-\langle\psi_L^-|).
\end{equation}

On the other hand, the probability conservation requires that $\langle I|P(t)\rangle=\langle I|e^{-\mathcal Ht}|P(0)\rangle=\langle I|P(0)\rangle=1$, where $\langle I|=\langle \psi^+_L|$. If the (right) even-parity steady state $|G_R^e\rangle$ is chosen as the initial state $|P(0)\rangle=|G^e_R\rangle$, the relation above requires
\begin{eqnarray}
\langle I|G_R^e\rangle=\frac{1}{2}(\mathcal N^e\langle G^e_L|+\mathcal N^o\langle G^o_L|)|G^e_R\rangle=\frac{1}{2}\mathcal N^e\langle G^e_L|G^e_R\rangle=1.
\end{eqnarray}
This enforces $\mathcal N^e=2$ to guarantee $\langle G^e_L|G^e_R\rangle=1$. The same thing happens for the odd-parity steady state, i.e., $\mathcal N^o=2$. \\


On the other hand, for a closed chain, the boundary condition of fermionic degrees of freedom is periodic (PBC) or antiperiodic (APBC) depending on parity sector, where the probability conservation is separately fulfilled in each sector. This suggests that $\langle G_L^e|$ and $\langle G_L^o|$ are respectively the left steady sate of even and odd sector. The Hamiltonian for a closed chain is obtained by adding a boundary term to the OBC Hamiltonian: $\mathcal H_{\text{closed}}=\mathcal H_{\text{open}}+h_{\text{boundary}}$
where $h_{\text{boundary}}=h_{L,L+1}$. In the odd-parity sector with PBC, $c_{L+1}=c_1$, $h_{\text{boundary}}$ is minimised by a wavefunction in form of $\langle \text{vac}|(1\pm c_1)(1\pm c_L)f(c_2,\cdots, c_{L})$. Let $O_{a,b}^\pm=\Pi_{j=a}^b(1\pm c_j)$ (with $b>a$),
\begin{eqnarray}
O^\pm_{L,1}=O^\pm_{L,2}\pm O^\pm_{L,2}c_1= O^\pm_{L,2}\pm c_1O^\mp_{L,2}.
\end{eqnarray}
This relation leads to 
\begin{eqnarray}
O_{L,1}^+-O_{L,1}^-=(1+c_1)(1+c_L)\Pi_{j=2}^L(1+c_j)-(1-c_1)(1-c_L)\Pi_{j=2}^L(1-c_j).
\end{eqnarray}
Therefore, the PBC case is minimised by $\langle G_L^o|=\langle \text{vac}|(O_{L,1}^+-O_{L,1}^-)$. In the even sector with APBC, $c_{L+1}=-c_1$, $h_\text{boundary}$ is minimised by a wavefunction in form of $\langle \text{vac}|(1\pm c_1)(1\mp c_L)f(c_2,\cdots, c_{L})$, which can be verified in the same way as discussed above.  It is straightforward to check that
\begin{eqnarray}
O_{L,1}^++O_{L,1}^-=(1-c_1)(1+c_L)\Pi_{j=2}^L(1+c_j)+(1+c_1)(1-c_L)\Pi_{j=2}^L(1-c_j).
\end{eqnarray}
Hence, $\langle G_L^e|=\langle \text{vac}|(O_{L,1}^++O_{L,1}^-)$ is the steady sate for the APBC case.

\section{Dynamics of observables}

In this section, we explore the expectation value of local particle densities $\rho_i(t)$. The main results are Eq.~\eqref{eq:steady_density} and~\eqref{eq:dynamics_Sup}. For a given initial sate $|P(0)\rangle$, the expectation value at time $t$ is calculated by
\begin{eqnarray}\label{eq:expectation_value}
\rho_i(t)=\langle I|\hat\rho_i e^{-\mathcal Ht}|P(0)\rangle
\end{eqnarray}
where $\langle I|=\langle G_L^e|+\langle G_L^o|$. Depending on the parity of the initial state, it can be decomposed to all possible excited states upon the steady state with the same parity. Only excitations with an even number of particles is allowed due the parity conservation. The initial state can be decomposed as below
\begin{eqnarray}\label{eq:projection}
|P(0)\rangle=&& a^{(0)}|G^\tau_R\rangle
+\sum_{2\leq 2k\leq L}\sum_{(n_1,\cdots,n_{2k})\in P_{2k}} a^{(2k)}_{n_1,\cdots, n_{2k}}|n_1,\cdots,n_{2k}\rangle_\tau.
\end{eqnarray}
Here, the first term stands for the projection on the steady state, where $\tau=e/o$ is chosen to match with the parity of $|P(0)\rangle$, and  $|G^{e(o)}_R\rangle$ is the even-parity (odd-parity) right steady state. The second term represents projections on excited states, with the excitation $(n_1,\cdots,n_{2k})$ chosen from $P_{2k}$, a set encompassing all possible choices for the excitation of $2k$ quasi-particles.

When the zero mode is not involved into the excitation, $|n_1,\cdots,n_{2k}\rangle_{\tau}=\bar{\alpha}^\dagger_{n_1}\cdots\bar{\alpha}^\dagger_{n_{2k}}|G_R^\tau\rangle$ with subindices ordered by following $\varepsilon_{n_1}<\varepsilon_{n_2}<\cdots<\varepsilon_{n_{2k}}$, and $a^{(2k)}_{n_1,\cdots, n_{2k}}=\langle \bar{n}_1,\cdots, \bar{n}_{2k}|_\tau P(0)\rangle$
with $\langle \bar{n}_1,\cdots, \bar{n}_{2k}|_{\tau}=\langle G_L^\eta|\alpha_{n_{2k}}\cdots\alpha_{n_1}$. It is a little bit subtle when the zero mode is involved. If the zero mode in the steady state is not occupied, the projection works as before. However, if the steady state with correct parity is already occupied by the zero mode,  the excitation with one more zero mode is ruled out due to the Pauli exclusion principle. Instead, the parity conservation and steady state degeneracy allow the  ``excitation" composed of the annihilation of the zero model plus excitation of an odd number of other modes.  Another subtle point is that, in general, it is not an easy task to distinguish which steady state is occupied by the zero mode. Fortunately, it is not necessary to know the answer, because only the correct projection can survive (the other one automatically vanishes due to Pauli exclusion principle), even if two contributions, adding and killing the zero mode, are simultaneously considered.  Therefore, for these containing zero mode $\varepsilon_1=0$ or $\varepsilon_{L+1}=0$, we consider the following two terms:
\begin{eqnarray}
|1,\cdots,n_{2k}\rangle_{\tau}=\bar{\alpha}^\dagger_{1}\cdots\bar{\alpha}^\dagger_{n_{2k}}|G_R^\tau\rangle,\qquad \text{with}\quad a^{(2k)}_{1,n_2,\cdots,n_{2k}}=\langle G_L^\tau|\alpha_{n_{2k}}\cdots\alpha_{1}|P(0)\rangle,\nonumber\\
|L+1,\cdots,n_{2k}\rangle_{\tau}=\alpha_{1}\cdots\bar{\alpha}^\dagger_{n_{2k}}|G_R^\tau\rangle ,\qquad \text{with}\quad a^{(2k)}_{L+1,n_2,\cdots,n_{2k}}=\langle G_L^\tau|\alpha_{n_{2k}}\cdots\bar{\alpha}^\dagger_{1}|P(0)\rangle.
\end{eqnarray}

Since $\hat\rho_i$ takes a bilinear form, it takes the following form under eigenbasis representation
\begin{eqnarray}\label{eq:bilinear}
\hat\rho_i=\sum_{i,j=1}^L(\lambda^{(i)}_{1,ij}\bar{\alpha}_i^\dagger\alpha_j+ \lambda^{(i)}_{2,ij}\alpha_i\bar{\alpha}_j^\dagger+\lambda^{(i)}_{3,ij}\bar{\alpha}_i^\dagger\bar{\alpha}^\dagger_j+ \lambda^{(i)}_{4,ij}\alpha_i\alpha_j),
\end{eqnarray}
where the coefficients are functions of eigenstates (the diagonalization matrix)
\begin{eqnarray}
\lambda^{(i)}_{1,jl}=T_{ji}(T^{-1})_{il},\quad 
\lambda^{(i)}_{2,jl}=T_{j+L,i}(T^{-1}_{i,l+L}),\quad
\lambda^{(i)}_{3,jl}=T_{ji}(T^{-1})_{i,l+L},\quad
\lambda^{(i)}_{4,jl}=T_{j+L,i}(T^{-1})_{il}.
\end{eqnarray}
By plugging Eq.~\eqref{eq:projection} into Eq.~\eqref{eq:expectation_value},
 the expectation value becomes
\begin{eqnarray}\label{eq:expectation_OBC}
\rho_i(t)=a^{(0)}\langle I|\hat{\rho}_i|G^\tau_R\rangle +\underbrace{\sum_{1\leq n_1<n_2\leq L}a_{n_1,n_2}^{(2)}e^{-(\varepsilon_{n_1}+\varepsilon_{n_2}) t}\langle I|\hat\rho_i\bar{\alpha}_{n_1}^\dagger\bar{\alpha}_{n_2}^\dagger|G^\tau_R\rangle}_\text{$\delta\rho_i^{(1)}(t)$}+
\underbrace{\sum_{n_2=2}^La_{L+1,n_2}^{(2)}e^{-\varepsilon_{n_2}t}\langle I|\hat\rho_i\alpha_1\bar{\alpha}_{n_2}^\dagger|G^\tau_R\rangle}_\text{$\delta\rho_i^{(2)}(t)$}.
\end{eqnarray}
Here, the first term is the expectation value of the observable in the steady sate, the second and the third together constitute the dynamical part.

\textbf{Steady-state expectation}--- If the initial state is assigned to an empty chain, which has an even parity, the steady state projection will be $a^{(0)}=\langle G^{e}_L|0\rangle=1$.  Although it is not straightforward to test whether the steady state is occupied by the zero mode, one expects that the steady-state value should be independent of the parity of steady state (equivalent to say whether the zero mode is empty or occupied), especially for a large enough system. This means that no matter starting with an empty state or a state filled with one particle, we should end up with the same steady-state expectation. Indeed, in the following, we can show that one will arrive at the same result by assuming the zero mode is empty or occupied. If the zero mode is empty, i.e., $\alpha_1|G^e_R\rangle=0$, 
\begin{eqnarray}\label{eq:steady_state}
\rho_{i,\text{st}}=\langle I|\hat{\rho}_i|G^e_R\rangle=\sum_{n=1}^L\lambda^{(i)}_{2,nn}=\sum_{n=1}^LT_{n+L,i}(T^{-1}_{i,n+L});
\end{eqnarray}
Otherwise ( if $\bar{\alpha}^\dagger_1|G^e_R\rangle=0$),
\begin{eqnarray}\label{eq:steady_state1}
\rho_{i,\text{st}}=\lambda^{(i)}_{1,11}+\sum_{n=2}^L\lambda^{(i)}_{2,nn}.
\end{eqnarray}
These two cases will agree with each other if $\lambda^{(i)}_{1,11}=\lambda^{(i)}_{2,11}$. Specifically, $\lambda^{(i)}_{1,11}=T_{1,i}(T^{-1})_{i,1}$ with
\begin{eqnarray}
T_{1,i}=F_{1k}M_{kl}(V^\dagger)_{li}=\sum_{l=1}^{2L}\frac{1}{\sqrt{2}}(V^\dagger)_{li}(\bra{\bar{u}_{1}}l\rangle-i\bra{\bar{u}_{2}}l\rangle),\nonumber\\
(T^{-1})_{i,1}=V_{il}(M^{-1})_{lk}(F^{-1})_{k1}=\sum_{l=1}^{2L}\frac{1}{\sqrt{2}}V_{il}(\bra{l}u_1\rangle+i\bra{l}u_2\rangle).
\end{eqnarray}
Therefore,
\begin{eqnarray}
\lambda^{(i)}_{1,11}=\sum_{l,m=1}^{2L}\frac{1}{2}(V^\dagger)_{li}V_{im}(\langle\bar{u}_{1}\ket{l}-i\langle\bar{u}_{2}\ket{l})(\langle m\ket{u_{1}}+i\langle m\ket{u_{2}}).
\end{eqnarray}
In a similar way, it is readily to derive that 
\begin{eqnarray}
\lambda^{(i)}_{2,11}=T_{1+L,i}(T^{-1})_{i,L+1}=\sum_{l,m=1}^{2L}\frac{1}{2}(V^\dagger)_{li}V_{im}(\langle\bar{u}_{1}\ket{l}+i\langle\bar{u}_{2}\ket{l})(\langle m\ket{u_{1}}-i\langle m\ket{u_{2}}).
\end{eqnarray}
Due to $|u_{1(2)}\rangle=\langle\bar{u}_{1(2)}|$, the expected relation $\lambda_{1,11}=\lambda_{2,11}$ is verified, thus the value of steady-state expectation is independent of whether the zero mode in the steady state is occupied. The steady state density, either Eq.~\eqref{eq:steady_state}  or \eqref{eq:steady_state1}, can be rewritten in terms of the eigenstates:
\begin{equation}\label{eq:steady_density}
\boxed{\rho_{i,st}=\sum_{n=1}^L\langle\bar\psi_{n+L}|\ket{i}\bra{i}|\psi_{n+L}\rangle }.
\end{equation}

\textbf{Dynamical part}---To get the dynamical properties of the local particle density expectation value, it is necessary to determine the projection coefficients of the excited states. Since we are interested in an observable of the bilinear form, it is enough to consider excited states with two quasi-particles. As mentioned above, the subtlety of zero mode occupation complicates the calculation, where one needs to add two possibilities together and the correct one will be selected automatically. First, we consider the absence of an zero mode, in which all excitations, including the zero mode, can be treated on an equal footing.  By utilizing $\alpha_n=\sum_{j=1}^L T_{n,j}c_j+T_{n,j+L}c^\dagger_j$, one obtains
\begin{eqnarray}
a^{(2)}_{n_1,n_2}&&=\langle G^e_L|\alpha_{n_2}\alpha_{n_1}|0\rangle=\sum_{j,l=1}^L T_{n_2,j}T_{n_1,l+L}\delta_{jl}+T_{n_2,j+L}T_{n_1,l+L}(1-\delta_{jl})\text{sgn}(l-j)\nonumber\\
&&=\sum_{j=1}^L T_{n_2,j}T_{n_1,j+L}+\sum_{l=1}^L\sum_{j<l}(T_{n_2,j+L}T_{n_1,l+L}-T_{n_2,l+L}T_{n_1,j+L}),
\end{eqnarray}
where we have used
\begin{eqnarray}
\langle G^e_L|c_i^\dagger c_j|0\rangle=\langle G_L^e|c_ic_j|0\rangle=0,\qquad
\langle G^e_L|c^\dagger_ic^\dagger_j|0\rangle=\text{sgn}(j-i)(1-\delta_{ij}),\quad
\langle G^e_L|c_ic_j^\dagger|0\rangle=\delta_{ij}.
\end{eqnarray}
It is worth stressing again that the ordering of fermionic operators in constructing the steady state in Eq.~\eqref{eq:steady_state_construction} is important to obtain correct projection coefficients.
From the definition, $a^{(2)}_{n_1,n_2}$ is antisymmetric with respect to the subscripts, i.e., $a^{(2)}_{n_1,n_2}=-a^{(2)}_{n_2,n_1}$. The second part in the expression automatically fulfills this requirement, while it is not so obvious for the first part. In fact, this relation can be shown as below
\begin{eqnarray}
\sum_{j=1}^L T_{n_2,j}T_{n_1,j+L}+ \sum_{j=1}^L T_{n_1,j}T_{n_2,j+L}=\sum_{j=1}^L[T_{n_2,j}(T^{-1})_{j,n_1+L}+T_{n_2,j+L}(T^{-1})_{j+L,n_1+L}]=(TT^{-1})_{n_2,n_1+L}=0.
\end{eqnarray}
On the other hand, the time-dependent part in Eq.~\eqref{eq:expectation_OBC} has a component
 $\langle I|\hat\rho_i\bar{\alpha}_{n_1}^\dagger\bar{\alpha}_{n_2}^\dagger|G^e_R\rangle=\sum_{i,j}\lambda^{(i)}_{4,ij}(\delta_{in_2}\delta_{jn_1}-\delta_{in_1}\delta_{jn_2})$. By substituting this relation into  Eq.~\eqref{eq:expectation_OBC}, the dynamic part of the local particle number expectation gains a contribution as below
\begin{eqnarray}\label{eq:particle_dynamics1}
\delta \rho^{(1)}_i(t)=\sum_{n_1<n_2}a^{(2)}_{n_1,n_2}e^{-(\varepsilon_{n_1}+\varepsilon_{n_2})t}(\lambda^{(i)}_{4,n_2n_1}-\lambda^{(i)}_{4,n_1n_2}).
\end{eqnarray}

Next, we consider the case that the steady-state is already occupied by the zero mode. Following the same logic, it is easy to obtain
\begin{eqnarray}
a^{(2)}_{L+1,n_2}=\langle G^e_L|\alpha_{n_2}\bar{\alpha}^\dagger_{1}|0\rangle=(1-\delta_{n_2,1})[\sum_{j=1}^L T_{n_2,j}T_{L+1,j+L}+\sum_{l=1}^L\sum_{j<l}(T_{n_2,j+L}T_{L+1,l+L}-T_{n_2,l+L}T_{L+1,j+L})].
\end{eqnarray}
Moreover, the corresponding part involving the zero mode is $\langle G_L^e|\hat{\rho}_i\alpha_1\bar{\alpha}_{n_2}^\dagger|G^e_R\rangle=(1-\delta_{n_2,1})(\lambda^{(i)}_{2,n_21}-\lambda^{(i)}_{1,1n_2})$.  Hence, the other possibility could generate a contribution
\begin{eqnarray}\label{eq:particle_dynamics2}
\delta\rho_i^{(2)}(t)=\sum_{n_2=2}^La^{(2)}_{L+1,n_2}(\lambda^{(i)}_{2,n_21}-\lambda^{(i)}_{1,1n_2}) e^{-\varepsilon_{n_2}t}.
\end{eqnarray}
In sum, the dynamic part is the addition of Eq.~\eqref{eq:particle_dynamics1} and Eq.~\eqref{eq:particle_dynamics2}: 
\begin{eqnarray}\label{eq:dynamics_Sup}
\boxed{\delta\rho_i(t)=\sum_{1\leq n_1<n_2\leq L}\Gamma^{(i)}_{n_1,n_2}e^{-(\varepsilon_{n_1}+\varepsilon_{n_2})t}
},
\end{eqnarray}
with
\begin{eqnarray}
&&\Gamma_{1,n}^{(i)}=a^{(2)}_{1,n}(\lambda^{(i)}_{4,n1}-\lambda^{(i)}_{4,1n})+a^{(2)}_{L+1,n}(\lambda^{(i)}_{2,n1}-\lambda^{(i)}_{1,1n}),\qquad 2\leq n\leq L,\nonumber\\
&&\Gamma_{n_1,n_2}^{(i)}=a^{(2)}_{n_1,n_2}(\lambda^{(i)}_{4,n_2n_1}-\lambda^{(i)}_{4,n_1n_2}), \qquad 2\leq n_1,n_2\leq L.
\end{eqnarray}

Due to the NHSE, the wavefunction profile for continuum bands is roughly sketched by $\bra{j}\ket{\psi_n}\sim |\beta|^j$; while the zero-mode wavefunction is the superposition of two isolated zero modes, its shape ($\bra{j}\ket{\psi_1}$) is asymptotically given by $C_+\beta_+^j+C_- \beta_-^j$ with $|\ln |\beta_\pm||^{-1}$ being the localization length of the zero mode, and $C_\pm$ being combination coefficients.
Therefore, the location dependence is roughly approximated as following: for $\Gamma_{1,n}^{(i)}$,
\begin{eqnarray}
\lambda^{(i)}_{4,n1}-\lambda^{(i)}_{4,1n}
\sim |\beta|^i(C_+\beta_+^i+C_- \beta_-^i),\qquad
\lambda^{(i)}_{2,n1}-\lambda^{(i)}_{1,1n}
\sim |\beta|^i(C_+\beta_+^i+C_- \beta_-^i);
\end{eqnarray}
for $\Gamma_{n_1,n_2}^{(i)}$ with $n_1,n_2\geq 1$,
\begin{eqnarray}
\lambda^{(i)}_{4,n_2n_1}-\lambda^{(i)}_{4,n_1n_2}
\sim |\beta|^{2i}.
\end{eqnarray}
As a special case, for the parameters chosen for Fig.2 (c),(d) in main text, $s=\delta_1>\delta_2>0$, Eq.~\eqref{eq:beta} and ~\eqref{eq:MZM_localization} tell us
\begin{eqnarray}
\beta_+=|\frac{\delta_1-\delta_2}{\delta_1+\delta_2}|<1,\qquad \beta_-=1,
\end{eqnarray}
which indicates that the magnitude of location-dependent coefficient $\Gamma_{n_1,n_2}^{(i)}$ decreases from left boundary to right boundary, agreeing with the observation in Fig.2 (d) in main text.

\subsection{Another approach for the dynamical evolution}

In this section, we provide another approach of studying the time-dependent particle density expectation value. Consider a general correlator
\begin{eqnarray}
C_{ij}=\langle\Psi^\dagger_i\Psi_j\rangle=\langle I|\Psi^\dagger_i\Psi_j e^{-\mathcal Ht}|P(0)\rangle.
\end{eqnarray}
The time-evolution equation for the correlator is
\begin{eqnarray}\label{eq:correlator}
\frac{d}{dt} C_{ij}=\langle I|\Psi^\dagger_i\Psi_j(-\mathcal H) e^{-\mathcal Ht}|P(0)\rangle=\langle I|[\mathcal H,\Psi^\dagger_i\Psi_j]|P(t)\rangle
\end{eqnarray}
where $\langle I|\mathcal H=0$ is applied. The commutator is calculated as below
\begin{eqnarray}\label{eq:communitator}
[\mathcal H,\Psi^\dagger_i\Psi_j]&&=\frac{1}{2}H_{mn}[\Psi^\dagger_m\Psi_n,\Psi^\dagger_i\Psi_j]\nonumber\\
&&=\frac{1}{2}H_{mn}[\delta_{ni}\Psi^\dagger_m\Psi_j-(\Sigma_x)_{nj}\Psi_m^\dagger\Psi_i^\dagger+(\Sigma_x)_{mi}\Psi_j\Psi_n-\delta_{mj}\Psi_i^\dagger\Psi_n]\nonumber\\
&&=\frac{1}{2}[H^T_{im}\Psi_m^\dagger\Psi_j-(\Sigma_xH\Sigma_x)_{lj}\Psi_l\Psi_i^\dagger+(\Sigma_xH\Sigma_x)_{il}\Psi_j\Psi_l^\dagger-\Psi^\dagger_i\Psi_n H^T_{nj}]\nonumber\\
&&=H^T_{il}\Psi^\dagger_l\Psi_j-\Psi^\dagger_i\Psi_l H^T_{lj}.
\end{eqnarray}
Here, the anticommutator $\{\Psi_m^\dagger,\Psi_n\}=\delta_{mn}$ and $\{\Psi_m^\dagger,\Psi_n^\dagger\}=\{\Psi_m,\Psi_n\}=(\Sigma_x)_{mn}$ are utilized in the second line; The particle-hole symmetry $\Psi_m^\dagger=(\Sigma_x)_{mn}\Psi_n$, $\Psi_m=\Psi_n^\dagger(\Sigma_x)_{mn}$, and $\Sigma_x H\Sigma_x=-H^T$ are applied from the second line to the third line. By plugging Eq.\eqref{eq:communitator} back to Eq.\eqref{eq:correlator}, the equation can be packed into a matrix equation
\begin{eqnarray}
\frac{d}{dt} C=[H^T, C].
\end{eqnarray}
The time-dependent form of the correlator matrix reads
\begin{eqnarray}\label{eq:time_dependent_correlator}
C(t)=e^{H^T t}C(0)e^{-H^T t}.
\end{eqnarray}
For an empty initial state, the initial correlator takes the following block form
\begin{eqnarray}
C(0)=\left(\begin{array}{cc}
    \boldsymbol 0 & X  \\
     \boldsymbol 0& \mathbbm{1}_{L\times L} 
\end{array}\right)
\end{eqnarray}
where 
\begin{eqnarray}
X_{ij}= \langle I|c^\dagger_ic^\dagger_j|0\rangle=\text{sgn}(j-i)(1-\delta_{ij}).
\end{eqnarray}
In this approach, the time-dependent particle number can be extracted from the diagonal component of the correlator $C(t)$ which  evolves according to Eq.~\eqref{eq:time_dependent_correlator}. By directly implementing the matrix evolution form, we exactly obtained the same results as before.

\begin{figure}
\begin{tabular}{cc}
     \includegraphics[width=0.7\linewidth]{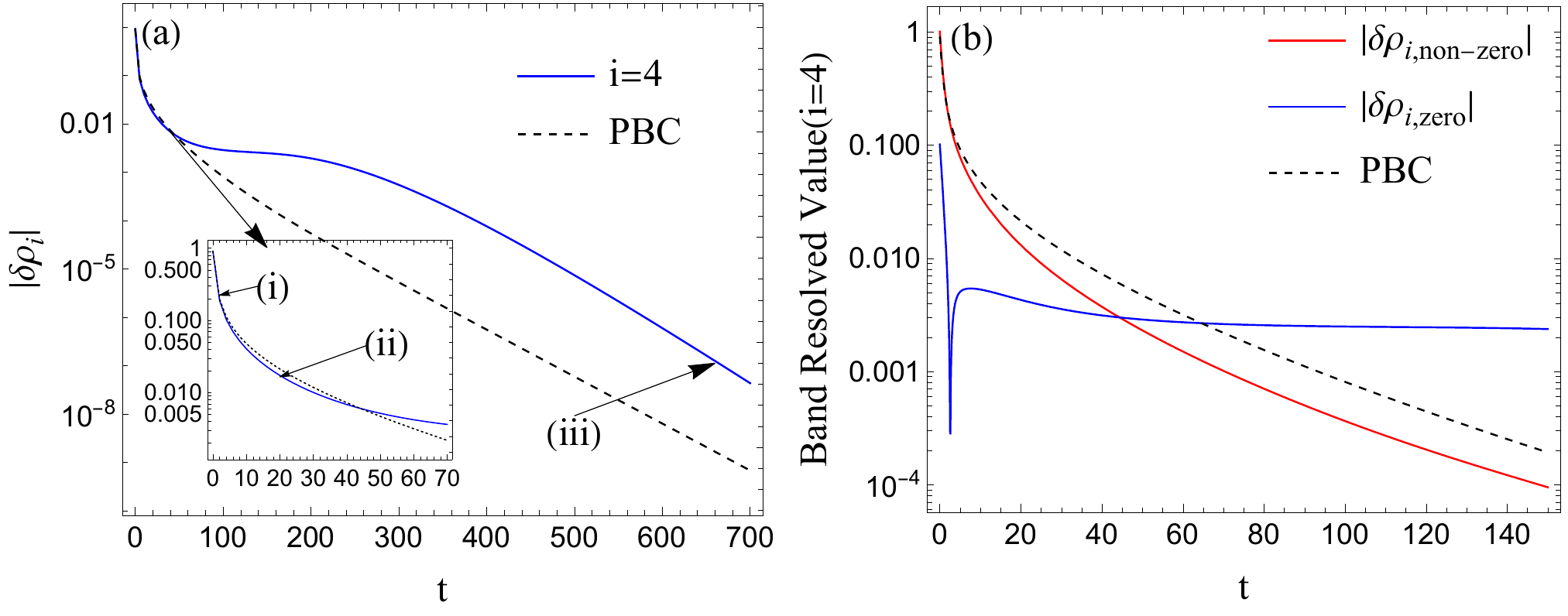} \\
      \includegraphics[width=0.7\linewidth]{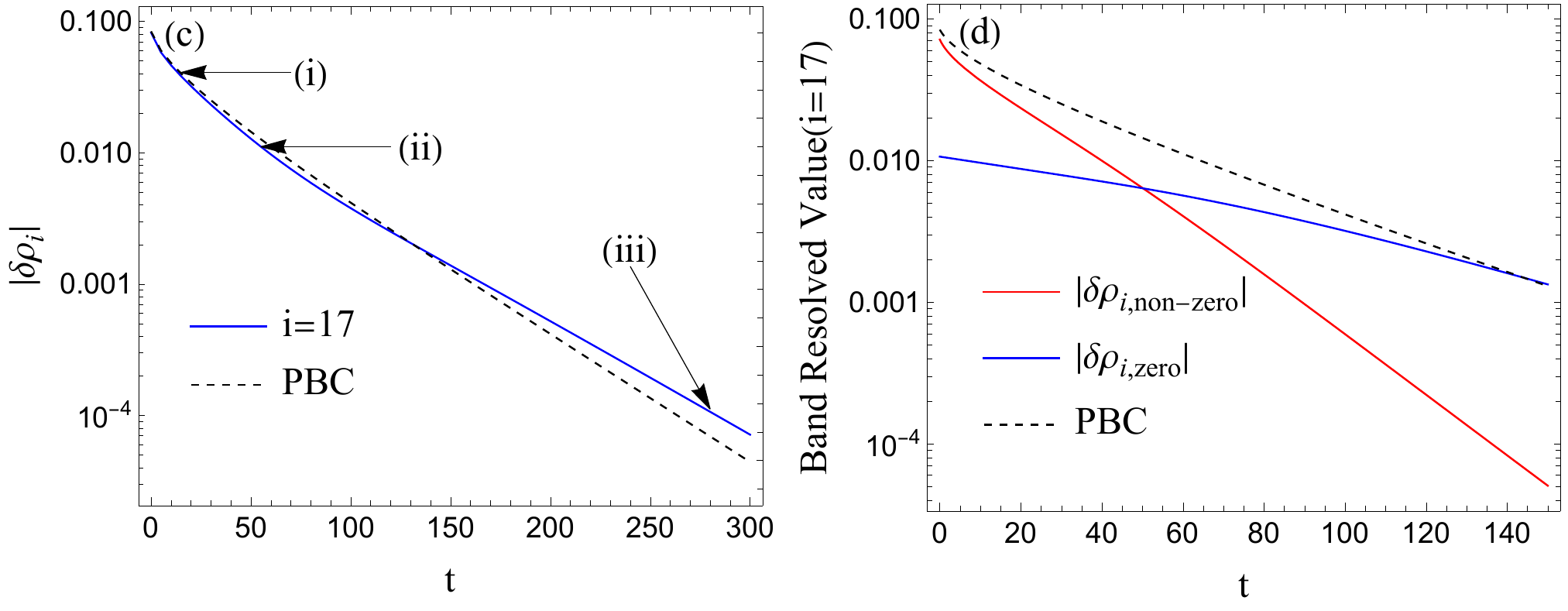}
\end{tabular}
 \caption{Examples with the middle stage (ii), dominating by NHSE, shows up. (a),(c) plot the dynamic part of the particle number expectation value at a given site. (b),(d) show the band-resolved contributions, distinguished by whether it contains the zero mode. Parameters are taken as the following,  (a), (b):$s=0.3$, $\delta_1=0.295$, $\delta_2=0.07$. (c),(d):$s=0.3$, $\delta_1=-0.295$, $\delta_2=0.05$.}\label{fig:Middle_stage}
\end{figure}

\subsection{Middle stage in the damping process}

The time evolution process of the particle number expectation might be divided into three stage as the following. (i) 
It usually takes a period for a given position in the chain to feel the existence of the boundary. Therefore, the particle number at this position evolves by following the PBC trajectory within this period. After this period, the boundary effects, including non-Hermitian skin effect (NHSE) and the modulation by Majorana-like zero modes, start to emerge.  (ii) The effect of NHSE is manifested when the decay rate of $\delta\rho_i(t)$ is dominated by two nonzero eigenvalues (after deviating for PBC trajectory), especially by $2\Delta_{\text{OBC}}$ when all other higher modes gradually die out. (iii) The effect of the zero mode is featured with the decay rate dominated by a single nonzero eigenvalue, which finally shrinks to the band bottom with a gap (real part) $\Delta_{\text{OBC}}$. 

Since the stage (ii) has a much faster decay rate than stage (iii), the latter will eventually govern the behavior of $\delta \rho_i(t)$ with the rate $\Delta_{\text{OBC}}$. It is even possible that the stage (ii) does not appear at all if the proportion of the decay channels with zero modes override other channels immediately after OBC dynamics deviating from the PBC trajectory. In reality, this case is most likely to happen in the parameter space. Therefore, the stage (ii) could show up only if the contribution from two nonzero modes is much larger than that containing a zero mode at the deviation point.  To illustrate this, we divide the dynamics into two parts distinguished by whether containing the zero mode, $\delta\rho_i(t)=\delta\rho_{i,\text{zero}}(t)+\delta\rho_{i,\text{non-zero}}(t)$  where
\begin{eqnarray}
&&\delta\rho_{i,\text{zero}}(t)=\sum_{2\leq n\leq L}\Gamma^{(i)}_{1,n}e^{-\varepsilon_{n}t},\nonumber\\
&&\delta\rho_{i,\text{non-zero}}(t)=\sum_{2\leq n_1<n_2\leq L}\Gamma^{(i)}_{n_1,n_2}e^{-(\varepsilon_{n_1}+\varepsilon_{n_2})t}.
\end{eqnarray}
Fig.~\ref{fig:Middle_stage} demonstrates two examples where the stage (ii) appears. In (b) and (d), we find that the magnitude of the contribution without zero mode $|\delta\rho_{i,\text{non-zero}}|$ greatly exceeds the magnitude of the components with zero mode involved  $|\delta\rho_{i,\text{zero}}|$ when the OBC dynamics just departures from the PBC trajectory. This ensures that the stage (ii) will not be dominated by the stage (iii), as shown in (a), (c).

\bibliography{dirac,NHMajorana}